\documentclass[acmsmall,natbib]{acmart}
\setcopyright{acmcopyright}
\copyrightyear{2020}
\acmYear{2020}
\acmDOI{10.1145/0000000.0000000}

\acmJournal{PACMPL}
\acmVolume{4}
\acmNumber{OOPSLA}
\acmArticle{1}
\acmMonth{10}

\usepackage{natbib}

\usepackage{tikz}
\usepackage{amsmath}

\RequirePackage[frozencache]{minted}
\usepackage{epsfig}
\usepackage{comment}
\usepackage{graphicx}
\usepackage{caption}
\usepackage{subcaption}
\usepackage{listings}
\usepackage{graphicx}
\usepackage{pifont}
\usepackage{booktabs}
\usepackage{graphicx}
\usepackage{rotating, multirow}
\usepackage{makecell}
\usepackage{color}

\definecolor{bg}{rgb}{0.95,0.95,0.95}

\begin{document}
\bibliographystyle{plainnat}

\date{}

\newcommand{\systemclone}{\textsc{Mossad$_{\mbox{det}}$}}
\newcommand{\systemshatter}{\textsc{Mossad$_{\mbox{nondet}}$}}
\newcommand{\systemfull}{\textsc{Mossad}}
\newcommand{\systemname}{\textsc{Mossad}}
\newcommand{\systemjplag}{\textsc{Mossad-JPlag}}

\title{\systemname: Defeating Software Plagiarism Detection}

\author{Breanna Devore-McDonald}
\email{bdevorem@cs.umass.edu}
\author{Emery D. Berger}
\email{emery@cs.umass.edu}
\affiliation{%
  \department{CICS}
  \institution{University of Massachusetts Amherst}
  \streetaddress{140 Governors Drive}
  \city{Amherst}
  \state{Massachusetts}
  \postcode{01002}
  \country{USA}
}


\begin{abstract}
Automatic software plagiarism detection tools are widely used in
educational settings to ensure that submitted work was not
copied. These tools have grown in use together with the rise in
enrollments in computer science programs and the widespread
availability of code on-line. Educators rely on the robustness of
plagiarism detection tools; the working assumption is that the effort
required to evade detection is as high as that required to actually do
the assigned work.

This paper shows this is not the case. It presents an entirely
automatic program transformation approach, \systemname{}, that defeats
popular software plagiarism detection tools.
\systemname{} comprises a framework that couples techniques inspired by
genetic programming with domain-specific knowledge to effectively
undermine plagiarism detectors. \systemname{} is effective at
defeating four plagiarism detectors, including
Moss~\citep{schleimer2003winnowing} and
JPlag~\citep{prechelt2002finding}. \systemname{} is both fast and
effective: it can, in minutes, generate modified versions of programs
that are likely to escape detection. More insidiously, because of its
non-deterministic approach, \systemname{} can, from a single program,
generate \emph{dozens} of variants, which are classified as no more
suspicious than legitimate assignments. A detailed study
of \systemname{} across a corpus of real student assignments
demonstrates its efficacy at evading detection. A user study shows
that graduate student assistants consistently
rate \systemname{}-generated code as just as readable as authentic
student code. This work motivates the need for both research on more
robust plagiarism detection tools and greater integration of naturally
plagiarism-resistant methodologies like code review into computer
science education.

\end{abstract}

\maketitle

\section{Introduction}
\label{sec:introduction}

Plagiarism in programming courses is unfortunately widespread.
Universities report that 10--70\% of their students have cheated on
coding assignments~\citep{eab_2017,Baron.2017,bidgood_merrill_2017,murray_2010}.
Two factors combine to effectively incentivize software plagiarism:

\begin{itemize}

\item \textbf{Homework solutions are often available on-line.}

The rise of open source hosting sites and programming-oriented Q\&A
sites has had numerous benefits, but also offers temptations for
misuse.  Open source hosting sites like GitHub often contain full
assignment solutions and code, in addition to solutions to typical
undergraduate projects~\citep{jue_2014, mcmillan_2015}. Sites like
Stack Overflow offer programmers abundant resources for coding
assistance and tips, which can also provide solutions to coursework;
the FAQ for Stack Overflow specifically states that ``it is okay to
ask about homework.''~\citep{stackoverflow-faq}.

\item \textbf{Enrollment pressure in computer science makes manual inspection impractical.}

The rise in computer science enrollments in recent years is
well-documented. The CRA reported in 2017 that there was a 185\%
``surge'' in undergraduate program enrollments since
2006~\citep{Camp:2017:GCG:3095781.3084362}. This rise in enrollments
has led to increased class sizes and faculty workloads, and shortages
in the number of faculty, instructors, and teaching
assistants~\citep[pages~5--6]{national2018assessing}. These factors
combine to make manual inspection of assignments impractical in many
cases. In a survey of computer science faculty conducted by the
authors (Section~\ref{sec:survey_for_threat_model}), roughly half of
the respondents said they perform little to no manual inspection of
solutions. Respondents cited ever-growing class sizes (from 150 to
nearly 1,000 students) as a reason for this practice.

\end{itemize}

Plagiarism thus represents an excellent (short-term) cost-benefit
approach for students. Students can easily search for and copy and
paste solutions to assignments, minimizing their effort. The absence
of oversight due to large class sizes further reduces the risk of
detection~\citep{10.1145/3159450.3159490}.

Software plagiarism detection tools like
Moss~\citep{schleimer2003winnowing,aiken_2020} can change this
equation by raising the risk of detection, thus making plagiarism less
attractive.  For software plagiarism detection to be effective, the
cost of defeating it must be high, and the risk of detection must be
low. Ideally, the difficulty of evading detection must be as high as
the effort required to actually do the assignment itself; this feature is
precisely what educators rely on, and what designers of plagiarism
detectors intend. The author of one of the plagiarism detectors we
analyze here (Sherlock~\cite{joy1999plagiarism}) states that ``if
(students) can plagiarise and avoid detection by Sherlock (or JPlag or
MOSS) then they can program to a good standard
already.''~\cite{joy_2020}

Existing software plagiarism detectors work primarily by identifying
suspiciously-similar code segments via a combination of tokenization
and hashing or directly comparing
strings~\citep{schleimer2003winnowing,chen_francia_li_mckinnon_seker_2004,Liu:2006:GDS:1150402.1150522,prechelt2002finding,syntaxbased}.
For example, Moss's tokenization effectively undoes the effect of
trivial changes like modifying formatting, editing comments, and
renaming variables or functions. It also identifies chunks of code
that are structurally similar across pairs of assignments, and
computes a score corresponding to the percentage of
matches. (Section~\ref{sec:background} describes Moss's algorithms in
detail.) Moss's approach generally makes it difficult for students to
manually alter code in a way that would elude
detection~\citep{Bowyer1999ExperienceU}.

Accordingly, software plagiarism detectors---especially Moss--have
seen widespread and growing adoption. The move to online courses due
to the coronavirus pandemic has spurred a further
increase~\citep{aiken_2020}.  Moss has also recently been integrated
into Gradescope, a widely used suite of grading
tools~\citep{gradescope_2019, inproceedings, forbes_2018}.

\subsection{\systemname{}}

This paper presents \systemname{}, a program transformation framework
inspired by genetic programming~\citep{le2011genprog} that defeats
popular software plagiarism detection tools, including Moss and
JPlag. \systemname{} can automatically produce plagiarized
assignments that escape detection; that is, the resulting similarity
scores between the original and plagiarized version of the code are
indistinguishable from those of different (legitimate) assignments.

\systemname{} thus defies the conventional wisdom that defeating plagiarism detection
is difficult or requires significant programming ability. The
techniques that underlie \systemname{} could be implemented manually,
relying on only the most basic understanding of programming language
principles, letting them evade detection by both plagiarism detectors
and some degree of manual inspection. The results of this paper both
indicate the need for research into more robust plagiarism detection
techniques and highlight the fact that common practice needs to be
revisited to account for this kind of plagiarism.

Section~\ref{sec:background} first provides a detailed technical
description of Moss's algorithms, which are key to understanding the
attack that \systemname{} embodies. Section~\ref{sec:analysis}
presents our threat model, informed by surveys of faculty, and
performs a security analysis of Moss's algorithms;
Section~\ref{sec:mossad} then describes the implementation
of \systemname{} and how it leverages the discovered vulnerabilities.

Section~\ref{sec:evaluation} empirically demonstrates \systemname{}'s
real-world effectiveness by conducting all experiments on a suite of
actual classroom projects from multiple institutions. The evaluation
shows that \systemname{} can take a single input program and generate
a variant that Moss, JPlag~\citep{prechelt2002finding}, and
Sherlock~\citep{joy1999plagiarism} all consider no more similar than
legitimate programs. More insidiously, because of its
non-deterministic nature, \systemname{} can also
generate \emph{dozens} of variants from a single input, each of which
is considered unsuspicious. We conduct a user study of graduate
student assistants trained as teaching assistants and find that they
rate \systemname{}-generated code as just as readable as authentic
student code, making even manual detection
unlikely. Section~\ref{sec:obfuscation} demonstrates \systemfull{}'s
superiority to approaches like code obfuscation.

Section~\ref{sec:countermeasures} describes a \emph{partial}
countermeasure to \systemfull{}, applicable to compiled languages like
C and C++. Finally,
Section~\ref{sec:relatedwork} and \ref{sec:conclusion} discuss related
work and conclude.

\subsection{Contributions}

In sum, this paper makes the following contributions:

\begin{itemize}
  \item It presents a security analysis of Moss and identifies a key vulnerability;
  \item It describes \systemname{}, an automated program transformation framework that targets this vulnerability;
  \item It presents a detailed empirical evaluation demonstrating \systemname{}'s effectiveness on actual student code, highlighting its ability to produce low similarity scores and thus effectively eluding detection by Moss, JPlag, and Sherlock.
\end{itemize}

\subsection{Ethical Considerations}
\label{sec:ethics}
The goal of this work is to demonstrate that existing software
plagiarism detectors can be systematically and effectively
defeated, to raise questions to educators about
the efficacy and robustness of popular plagiarism detection tools, and to
motivate research on improving plagiarism detection systems to make them more
robust to attack. To ensure that this publication provides substantial advance
warning for instructors to take this threat into account, and
following standard practices for vulnerability disclosures, we plan to
embargo the release of \systemname{}.
However, in the interest of advancing science, we plan to
make \systemname{} available to researchers upon request.

\textbf{Responsible disclosure:} The authors disclosed the \systemname{}
attacks to the author of Moss, as well as the authors of Sherlock 
and JPlag, over 120 days prior to the publication of this work. As far as the 
authors are aware, there have not been any substantial changes to any of these
tools since this disclosure. In response to the disclosure, the authors had
conversations with both Alex Aiken (of Moss) and Mike Joy (of Sherlock)
regarding previous attacks against each of the respective tools and discussion
of \systemname{}~\cite{aiken_2018, joy_2020}. The authors did not receive a 
response from the authors of JPlag.

\textbf{Privacy concerns:} Throughout this study, we ensured that all of our experiments
meet community ethical standards. This work was deemed exempt by our
institutional review board (IRB), since the data we use presents no
more than minimal risk to our subjects. All student assignment data we
use was anonymized and assigned random identification numbers before
we received it; we therefore cannot infer the authors of the code from
any of our experiments. Email addresses obtained from
professors surveyed were solely used to validate their identity.

\section{Moss Background}
\label{sec:background}
Before delving into how \systemname{} operates, we first provide a
detailed description of the operation of the Moss software plagiarism
detector, which is necessary to understand why \systemname{}'s attack
is effective. While we also evaluate \systemname{} against another
plagiarism detector, JPlag~\cite{prechelt2002finding}
(Section~\ref{sec:jplag}), we focus our attention on Moss for the
following reasons:

\begin{itemize}

  \item \textbf{Moss is popular.} To the best of our knowledge, Moss
  is currently the most widely used software plagiarism detection
  system. Aiken reports that the Moss web service has roughly 300K
  current Moss accounts, with 50K--100K new Moss accounts per
  year~\cite{aiken_2018}, and it is now integrated with Gradescope.
  
  \item \textbf{Moss is general.} At the time of writing, Moss
  supports 24 different programming languages, making it to our
  knowledge by far the most widely-applicable plagiarism detector.
  
  \item \textbf{Moss represents the state of the art.} Empirically, we
  found Moss to be the most effective of the tools we tested. Its
  design is similar to that employed by other leading plagiarism
  detectors.
    
\end{itemize}

As Section~\ref{sec:introduction} explains, Moss is an automatic
system for detecting the similarity of software provided as a free web
service, and it is widely used to detect plagiarism of classroom
assignments. Moss measures the similarity between every combination of
two input source programs (that is, all $O(n^2)$ pairs) and computes
the similarity score as the percentage of lines matched between each
pair. Beyond the above-cited characteristics, Moss has the following
properties that make it attractive as a plagiarism detector:

\begin{itemize}
    \item \textbf{Whitespace and comment insensitivity}: Moss ignores
    changes in whitespace, capitalization, or text in comments.
    
    \item \textbf{Noise suppression}: Moss exhibits a low false positive
    rate by reporting only large chunks of copied code. For example,
    the appearance of individual tokens like \texttt{int} in two files
    would not be reported as instances of plagiarism.
    
    \item \textbf{Position independence}: Moss can find plagiarized
    code regardless of its placement in the program.
    
\end{itemize}

\subsection{Normalization}

To provide whitespace insensitivity, Moss performs a
\emph{normalization step} before it moves on to actually computing
similarity. Depending on the language of the input files, Moss
eliminates irrelevant features that should not distinguish documents,
including semantically-irrelevant whitespace or comments. This
normalization step also consists of renaming all identifiers to the
same value; defeating variable renaming as a means of evading
detection. The input files are then tokenized into a normal form
across all input languages, and then sent as input to Moss's
\emph{fingerprinting engine}.

\subsection{Fingerprinting}

Moss's fingerprinting engine ensures both noise suppression and
position independence. It operates on the normal form that the
previous step produces; the same engine is used for all supported
source languages. The overarching goal of the fingerprinting engine
is to prepare the inputs for efficient similarity assessment by 
transforming the source files into small sets of values called \textit{fingerprints}. The algorithm aims to produce the fingerprints
with enough identifying information as to not
sacrifice correctness when assessing similarity.

As a first step, Moss divides the normalized program into a 
sequence of overlapping $k$-grams, a contiguous substring 
of length $k$, where $k$ is an internally-defined \textit{noise threshold}.
Then, Moss computes a hash for each $k$-gram, resulting in a sequence of 
multiple hashes representing the input (the fingerprints are chosen from this
sequence). The value of $k$ is particularly crucial in the trade-off between
noise suppression (false positive rate) and sensitivity: large $k$ would 
increase confidence in similarity matches, since it represents more source 
code; however, large $k$ would not allow Moss to detect shorter code matches.
As a result, the size of $k$ has a direct impact on positional
independence, since Moss is unable to detect relocation of substrings with
a length shorter than $k$. Though the value of
$k$ is unknown to the user, it can be obtained straightforwardly though
a brute force attack. 

After generating the sequence of hashes of $k$-grams, Moss then performs 
its unique \emph{winnowing} algorithm, which creates fingerprints by 
selecting a subset of the hashes to represent the entire input. The first 
step of winnowing is to group the hashes into overlapping windows of 
length $w$, where $w$ is the difference between a fixed 
\emph{guarantee threshold} $t$ and the \emph{noise threshold} $k$, plus one. 
The guarantee threshold is the minimum substring length, in terms of
tokens, that is to be detected when compared for similarity (intuitively, 
$k$ must be no greater than $t$, otherwise substrings 
of $t$ length would not be detectable).  Like the noise threshold, 
the guarantee threshold is an unknown 
value specified in the Moss fingerprinting engine. 

Since the window size is 
defined as $t - k + 1$, each substring of size $t$
has at least one corresponding window such that each hash in the window 
would be
detected as similar, if another input file contained that substring.
As a result, to scale down the set of fingerprints for an input, the
next step of winnowing is to choose a single hash from each window
to be a fingerprint. The algorithm for this selection is straightforward:
Moss chooses the \emph{minimum} hash value of each window. If the hash has
already been chosen as a fingerprint in the preceding window, then it will 
not be chosen again in the following window (essentially skipping 
successive windows 
with the same minimum hash value). If the minimum hash value appears
more than once in a window, Moss chooses the rightmost one. This 
algorithm actually further scales the set of fingerprints: the 
minimum hash value of a window
is most likely the minimum hash value of the next window (recall that
the windows overlap, similar to the $k$-grams), and therefore results
in far fewer fingerprints than the total number of $w$-sized windows.

The engine keeps track of the positional information of the
fingerprints, so that the source code of matches can be retrieved
later for reporting purposes. The engine adds the fingerprints it
chooses for each file along with the positional information to a
database. Then, each file is fingerprinted a second time and queried
in the fingerprint database for positional information, and the engine
returns the set of all matching fingerprints across the input
files. Rather than outputting the raw fingerprints to the user, Moss
returns the source code found from querying the fingerprint
database, such that the code matches are easily viewable by a human user.
Moss users can also specify that they want Moss to exclude
boilerplate or template code, in which case the fingerprints of the
boilerplate code are ignored if matched.

\begin{figure}
	\begin{subfigure}{.45\textwidth} 
        \begin{minted}[bgcolor=bg,fontsize=\small]{c}
int hello = 0;
return hello;
        \end{minted}
	\caption{A simple C program with headers and functions removed.\label{fig:hashing:a}}
	\end{subfigure}
	\vspace{1em} 
        \hspace{1em}
	\begin{subfigure}{.45\textwidth} 
		\begin{minted}[bgcolor=bg,fontsize=\small]{text}
TYP_INT ID EQ NUM SEMI
RET ID SEMI
	    \end{minted}
		\caption{The example set of tokens after smart tokenization of the C code in Figure~\ref{fig:hashing:a}.\label{fig:hashing:b}}
	\end{subfigure}
	\begin{subfigure}{.45\textwidth}
		\begin{minted}[bgcolor=bg,fontsize=\small]{text}
30 15 56 83 71 10
	    \end{minted}
		\caption{Representative hashes of the tokens from Figure~\ref{fig:hashing:b}, assuming $k = 3$.\label{fig:hashing:c}}
	\end{subfigure}
                \hspace{1em}
	\begin{subfigure}{.45\textwidth}
		\begin{minted}[bgcolor=bg,fontsize=\small]{text}
(30 15 56)  (15 56 83)
(56 83 71)  (83 71 10)
	    \end{minted}
		\caption{Windows of hashes of length $w = t - k + 1 = 3$ \label{fig:hashing:d}}
	\end{subfigure}
	\begin{subfigure}{.45\textwidth}
		\begin{minted}[bgcolor=bg,fontsize=\small]{text}
15 56 10
	    \end{minted}
		\caption{Post-\emph{winnowing}: keep the smallest hash in each $k$-gram, dropping consecutive duplicates, producing the fingerprints of the program.\label{fig:hashing:e}}
	\end{subfigure}
  \caption{\textbf{Moss Algorithm.} An example run of performing Moss's smart tokenization and fingerprinting algorithms, assuming that the guarantee threshold $t$ is 5 tokens and the noise threshold $k$ is 3.}
\label{fig:hashing}
\end{figure}

\subsection{Example}

Figure~\ref{fig:hashing} provides an example of Moss's normalization 
and fingerprinting
algorithms in action. Figure~\ref{fig:hashing:a} presents an example
consisting of two lines of valid C code. We omit headers and function
definitions for brevity and clarity. Figure~\ref{fig:hashing:b}
presents an example set of tokens that could result from
smart-tokenizing that code. Figure~\ref{fig:hashing:c} shows potential
hashes of these tokens, assuming a window of length 3 (3-gram
model). Figure~\ref{fig:hashing:d} shows the first stage in the
winnowing algorithm, where the $k$-gram hashes are further 
grouped into windows of size $w = t - k + 1$, which is 3 in this case.
Finally, Figure~\ref{fig:hashing:e} shows the result of finishing
Moss's winnowing algorithm. As previously noted, winnowing
creates a smaller set of hashes to represent the entire document
by choosing the smallest hash from each window of size $w$.

\section{Moss Security Analysis}
\label{sec:analysis}

This section performs a security analysis of Moss. First, we describe
our threat model, which was informed by the results of two surveys of
computer science faculty. Second, we verify empirically that Moss
operates as described in Section~\ref{sec:background}, which was based
on the paper describing it~\cite{schleimer2003winnowing}. Finally, we
identify a key weakness of Moss's algorithms, which \systemname{}
exploits.

\subsection{Developing the Threat Model}
\label{sec:survey_for_threat_model}

To inform a threat model (that is, establishing the power and
limitations of the adversary), we conducted two surveys on two
distinct social media platforms to understand the grading practices in
computer science courses, including their usage of software plagiarism
detectors.

We first conducted a poll on Twitter asking about the degree of manual
inspection performed on code. This poll asked respondents to
optionally provide further information regarding class sizes and time
spent on manual code inspection. Of the respondents (N=128), 65\% said
that they manually inspect nearly 0\% of student assignments. The
additional comments to the poll (N=12) reported that most student code
does not get manually inspected; specifically, the code that
\emph{does} get inspected are assignments flagged with high
similarity scores from Moss or other plagiarism detectors. (We did not
ask but believe that even when manual inspection is performed, it is
not conducted across every $N^2$ pair of assignments.)

We conducted a second, more detailed survey using Google Forms
advertised on Facebook and Twitter. Of the respondents to this survey
of faculty members (N=50 responses, 35 unique institutions), 65\%
report that they manually inspect nearly all of their assignments and
19\% report a manual inspection rate of nearly 0\%. This result
strongly suggests that the populations sampled by the two polls are
different. Aggregating the results across the two surveys yields an
average of 52\% of the respondents inspecting nearly no student
solutions.

Of the respondents to the second survey, 58\% of those reporting high
manual inspection rates also reported that they used Moss to inform
their decision of which assignments to inspect by either sorting based
on Moss scores or finding cliques from the Moss output. On the other
hand, 85\% of respondents reporting low manual inspection rates
limit their spot checks to the high matches reported by Moss.
Informed by the results of these surveys, we developed the following
threat model:

\subsection*{Threat Model}
\label{sec:threat_model}

\begin{itemize} 

\item Students are incentivized to minimize the time spent on assignments
      while maximizing their grade. Therefore, students will not spend
      more time attempting to defeat the plagiarism detector than
      actually doing the assignment.
      
    \item Students have arbitrary access to peers' assignment
      solutions or solutions found online. Students may use these
      solutions as the basis for their plagiarized assignment
      solution.

    \item In addition, all solutions are also available to the
        instructor; attacks that depend on outsourcing (hiring someone
        to write an entirely new program that is not available online)
        are to our knowledge beyond the scope of existing plagiarism
        detection tools, since the original program is unavailable.
    
    \item Course staff relies primarily on software plagiarism
      detection, beyond the lightweight manual inspection described
      below.
      
    \item Course staff will only examine student code for spot
      checking or for those assignments that receive high similarity
      scores with another assignment. Students will thus aim to
      achieve the lowest similarity score possible in order to avoid potential
      detection, but must also take care in that the submitted code
      retains readability; that is, submitting clearly obfuscated code
      would be an unacceptable risk in the case of a spot check.
      
\end{itemize}

The implications of our threat model are that the greatest risk to
current practices is a system that can automatically and rapidly
transform one source program into a multitude of variants that each
result in low Moss scores, while maintaining readability (as we show
in Section~\ref{sec:evaluation}, this is exactly what \systemname{}
achieves).

\subsection{Investigating Moss's Implementation}

With our threat model in hand, we next evaluate
whether the current implementation of Moss reflects its
description in the paper describing it~\cite{schleimer2003winnowing}
or if it has materially changed, since an attack based on the
algorithm described in the paper could plausibly fail against a new
implementation.

As Moss's source code is not publicly available, we could not manually
inspect it. Instead, we perform a series of experiments to evaluate
its behavior. Our experiments involve repeatedly transforming
differently-sized regions of code via a series of program
transformations, including code rearrangement and identifier renaming,
and observing their impact on Moss's reports.

We quickly were able to verify that noise suppression behaved as
expected; that is, Moss's window is sufficiently large to exclude
individual lines like \texttt{int x;} as constituting plagiarism.

The next experiment was to examine if Moss's whitespace insensitivity
worked as described. To do this, we took two copies of the same
program and edited one with comments, and changed the names of
identifiers. We performed this experiment automatically for 30 files
in our dataset (discussed in Section~\ref{sec:datasets}) and compared
the files with Moss. This experiment confirmed that whitespace
insensitivity worked as expected: the Moss scores before and after the
alterations did not differ.

We then examined Moss's position independence. To do this, for the 30
files, we created copies of files and rearranged the functions. After
this, we compared the files with Moss to see if the similarity score
had changed. In most cases, the similarity score did not change; in
the few cases in which it did, it did not go below an 85\% match. We
expect the change in Moss score is a product of Moss's scoring method,
rather than its similarity-detection algorithm. All of the code was
still detected as similar, but the code matches were separated into
multiple sections. This division into sections appears to affect the
percentage of similarity computed, though not to a large degree.

Finally, we performed the same experiment, but at a finer grain. By
rearranging code \emph{within} a function, we could examine how fine
of a granularity Moss can detect plagiarism--that is, approximately
how many lines of code correspond to a window of tokens. We tested
this by initially rearranging a single line of code, and then
increasing the number of lines that we rearranged by one at each
iteration.

We found that Moss could not detect plagiarism when just one or two
consecutive lines of code were rearranged. However, Moss could
consistently detect rearranged plagiarism for 3 and 4 consecutive
lines of code. Regions of greater than 5 lines of code were always
detected. As a result, we infer that the number of tokens used as the
length of the hashing window in Moss corresponds to between 3 and 4
lines of code.  This result is not precise, as the number of tokens
per line in a file can vary, but we found this to be the
case consistently within our dataset.

\textbf{Result:} We find that Moss's current implementation closely
resembles the original description. We confirm that Moss performs as
expected: it correctly assigns high similarity scores to code that has
been altered with static transformations. Crucially, we observe
behavior that is consistent with the hashing and winnowing approach
described in the original paper; it is this aspect of Moss's algorithm
that we identify as vulnerable to attack.

\subsection{Key Observation: Hash Disruption}
\label{sec:hash_disruption}

Recall that Moss hashes program tokens as $k$-grams (that is, $k$ is
the window size for hashing tokens). As mentioned earlier, this value
is a key parameter for Moss: a window size must be tuned so it is
large enough to reduce noise (false positives), yet small enough to
detect meaningful code matches (true positives). 

We hypothesized that Moss's algorithm suffers from the following
vulnerability, which we refer to as the \emph{hash disruption
hypothesis}: \textbf{inserting a single token in a window of length $k$
should cause Moss to fail to detect any similarity.}

To validate this hypothesis, we needed the value of $k$, which is not
made public and which may be different for each programming language
Moss supports. As described above, we reverse engineered a rough value
for $k$ by inserting statements at different strides. With our
approximate $k$ in hand, we were able to verify that, as anticipated,
a single change within the window leads to a different hash, since
Moss no longer treats such code fragments as similar.

\begin{figure}
	\begin{subfigure}[!t]{.45\textwidth}
        \begin{minted}[bgcolor=bg,fontsize=\small]{c}
int hello = 0;
bool nothing = true;
return hello;
        \end{minted}
		\caption{To demonstrate Moss's vulnerability, we first add a benign, semantics-preserving addition to the code from Figure~\ref{fig:hashing:a}.\label{fig:hashing2:a}}
	\end{subfigure}
    \hspace{1em}
    \begin{subfigure}[!t]{.45\textwidth}
		\begin{minted}[bgcolor=bg,fontsize=\small]{text}
TYP_INT ID EQ NUM SEMI 
TYP_BOOL ID EQ BOOL SEMI
RET ID SEMI
	    \end{minted}
		\caption{The example set of tokens after smart tokenization.\label{fig:hashing2:b}}
    \end{subfigure}
    \begin{subfigure}{.45\textwidth}
		\begin{minted}[bgcolor=bg,fontsize=\small]{text}
30 15 56 12 45 39 97 62 80 71 9
	     \end{minted}
      \caption{Potential hashes of these tokens.\label{fig:hashing2:c} Notice that the hash value 83 (shown in Figure~\ref{fig:hashing:b}) has been eliminated.}
	\end{subfigure}
  \caption{\textbf{Deterministically defeating Moss via \emph{hash disruption}.} A demonstration of
  \emph{hash disruption}, which exposes a key vulnerability of Moss's fingerprinting algorithm. After a benign
  addition to a series of statements, one of the hashes created by Moss's
  fingerprinting algorithm (83) is no longer present. \systemname{} leverages this vulnerability to drastically reduce the effectiveness of Moss's matching algorithm.}
\label{fig:hashing2}
\end{figure}

Figure~\ref{fig:hashing2} illustrates the hash disruption
vulnerability in simplified form. For the purposes of illustration, we
omit the effect of winnowing, which only minimizes the number of
hashes that are actually compared across files. The hash disruption
vulnerability described strikes at the heart of Moss's algorithm;
winnowing actually exacerbates its impact.

Recall the code in Figure~\ref{fig:hashing:a}. To effect hash
disruption, we simply add an additional line of code between the two
original lines. The new line of code creates a Boolean
variable. Following the same method of tokenization as used in
Figure~\ref{fig:hashing}, Figure~\ref{fig:hashing2:b} presents a
tokenization of the new C code. Then, Figure~\ref{fig:hashing2:c}
presents a hash of the tokens, again using the same method of hashing
as used previously. This figure shows that there are now more hashes
than shown in Figure~\ref{fig:hashing:c}, which is as expected as
there are now more tokens to be hashed. Hash 83 from
Figure~\ref{fig:hashing:b} has been disrupted and is no longer present
in the new set of hashes, leading to a failed match.

\subsection{Other Plagiarism Detection Tools}
\label{sec:intro_othertools}

In addition to Moss, we aimed to collect 
all of the related systems described in Section~\ref{sec:related_work}. 
This effort was largely unsuccessful, with three exceptions: JPlag~\cite{prechelt2002finding}, Sherlock~\cite{joy1999plagiarism}, and
Fett~\cite{syntaxbased}. JPlag and Sherlock are the only tools, aside from Moss, our survey respondents reported using. Other related systems 
are described in Section~\ref{sec:related_work}.

The algorithms used by both JPlag and Sherlock are similar in effect
to Moss. Like Moss, JPlag and Sherlock detect software
similarity. Both also employ tokenization and normalization of the
input file. However, rather than relying on hashing and winnowing to
build the file's fingerprints, JPlag and Sherlock both use Running
Karp-Rabin Greedy String Tiling to identify code matches and produce a
similarity score. This approach has the same effect as Moss's core
algorithm (excluding winnowing). 

On the other hand, the algorithm used by Fett is unlike that of
Moss. Rather than relying on tokenization, Fett uses parse trees with
weighted nodes in order to determine what needs to be filtered. For
the scoring stage of the algorithm, unlike the hashing and
fingerprinting schemes of Moss, Fett uses the Smith-Waterman algorithm
to align and identify common subsequences, while penalizing
gaps~\cite{smith1981identification}. Although there are no ``fingerprints'' as employed
by Moss, JPlag, and Sherlock, and therefore no hashes to be
disrupted, \systemname{} has a disruptive effect on Fett's algorithm
nonetheless. \systemname{} disrupts Fett's sequences by regularly
placing important nodes (such as useless assignments and empty
conditionals) at regular intervals, leading to large gap penalties.

In section \ref{sec:othertools}, we evaluate the effectiveness of \systemname{}
on JPlag, Sherlock, and Fett. As we demonstrate, while \systemname{} targets 
Moss's fingerprinting algorithm, it in effect implicitly targets the 
algorithms used by these systems as well, as they are all subject to the same 
disruption vulnerability.

\section{\systemname{}}
\label{sec:mossad}

The hash disruption phenomenon that Section~\ref{sec:analysis} describes
forms the groundwork for defeating detection: foil Moss's algorithm by
systematically introducing benign (semantics-preserving) statements
within windows. This approach is the point of departure
for \systemname{}.

\begin{figure}[!t]
    \centering
    \includegraphics[scale=.3]{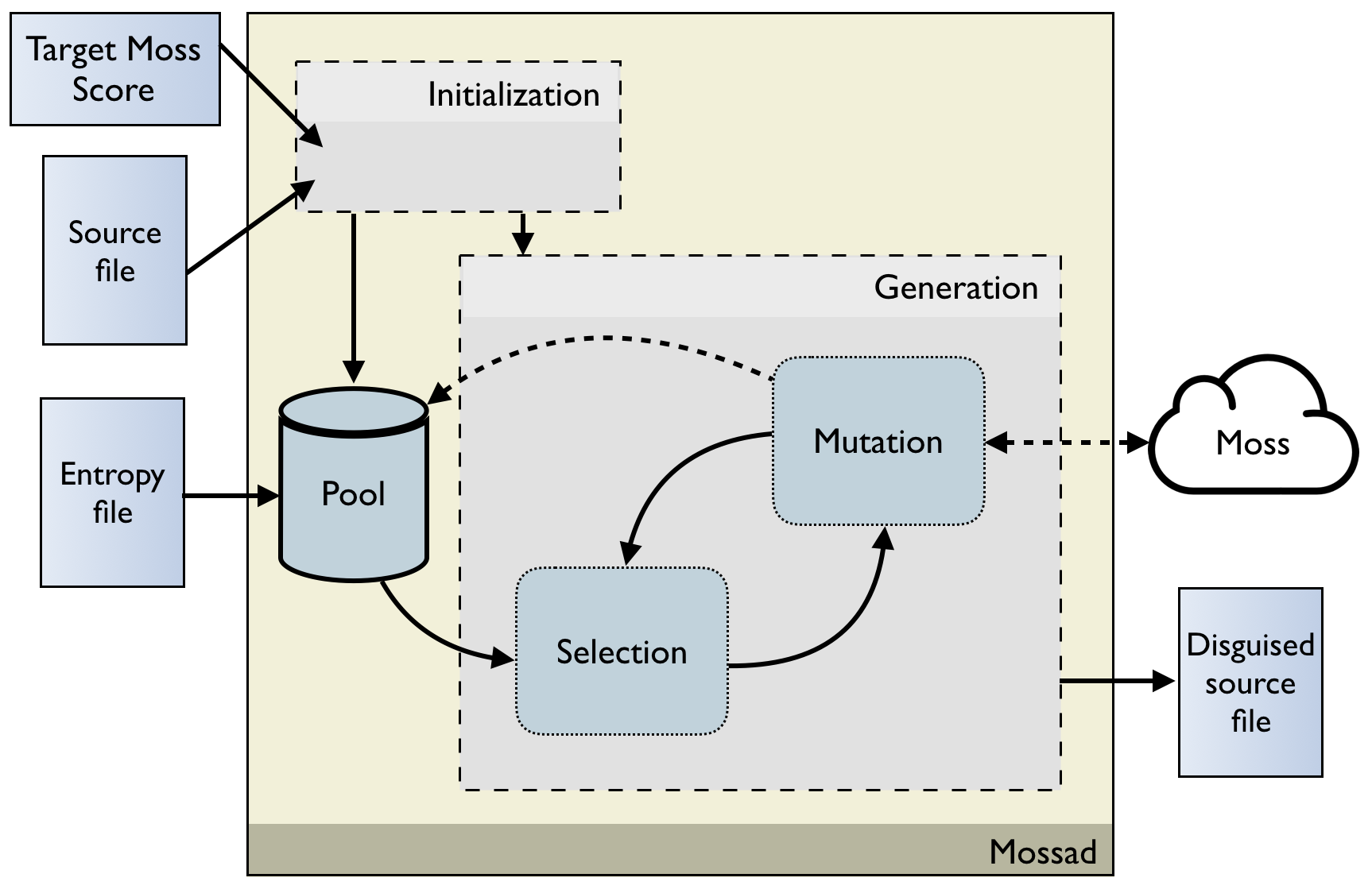}
    \caption{\textbf{\systemfull{} system overview.} \systemfull{} is a program transformation framework inspired by genetic programming to transform a file into semantics-preserving variants that defeat Moss's detection algorithms (Section~\ref{sec:mossad}).}
    \label{fig:system_diagram}
\end{figure}

\systemfull{} is a program transformation framework that fully
automates this approach. \systemfull{} presents the worst-case with respect to
the threat model developed in Section~\ref{sec:threat_model}: it is
not only entirely automated (requiring little to no effort) and fast
(taking minutes, as we describe in Section~\ref{sec:evaluation}), but
also can produce numerous variants that each escape detection by Moss,
even with quite low detection
thresholds. Figure~\ref{fig:system_diagram} presents an overview of
the system. Figure~\ref{fig:mossad_output} shows an example of running
\systemfull{} on an input function with a target Moss score of
10\%. Figure~\ref{fig:mossad_output:c} shows that Moss considers the original
and the \systemfull{}-processed variant to have no matches.

Given an input file that compiles without errors, \systemfull{} uses
techniques inspired by genetic programming to transform it into
semantically-equivalent variants. Given a user-provided target
similarity score (e.g., 25\%),
\systemfull{} will attempt to produce a variant that achieves a Moss
score less than or equal to the target score when compared to the
original input file. The current \systemname{} prototype only supports
C/C++ as input programming languages, but the technique described here
is directly applicable to any languages that can be statically
compiled.

\subsection{Initialization}

The first step of the \systemfull{} algorithm is the initialization
phase.  \systemfull{} compiles the input file (with optimization) to
an intermediate form. For C, \systemfull{} compiles the files with
Clang to object code. If compilation succeeds, the initialization
phase then creates a copy of the input source code, which becomes the
first variant. Initialization also seeds the \emph{pool}, which
initially consists of every line from the user input file, for use by
the generation phase.

\subsection{Generation}

The generation phase performs selections and mutations, and
reinitializes the pool for the next generation. The first module of
generation is selection. Selection chooses one line of code from the
pool and performs na\"ive semantic checks to speed up the subsequent
generation module. These checks include ensuring that the selected
line is a complete statement (e.g., ends with a semicolon for C) and
will not change the semantics of the program (e.g., is a print
statement). Once a selection is chosen, the selection engine sends the
chosen line to the next generation module: mutation.

\subsection{Mutation}

The mutation engine performs mutations by inserting the selection into
the file and ensuring that the original file's semantics are
preserved. To do this, the mutation engine first creates a temporary
variant (which is a copy of the current variant) to be
mutated. Mutation will randomly insert the selected line from
selection into the temporary variant and perform a sequence of checks.

In each generation, there are two outcomes for variants. The first is
a successful mutation, where the selection engine produces a
semantics-preserving statement. If the variant is not semantically
equivalent to the original (which is checked as described in
Section~\ref{sec:program_equiv}), then mutation discards the
temporary variant and restarts the selection module with the same pool
for regeneration.

Note that a mutation is considered successful whether or not the
target Moss score has been reached, since it almost always requires
multiple successful mutations to reach the target.

\begin{figure}[!t]
	\begin{subfigure}[!t]{.45\textwidth}
		\begin{minted}[bgcolor=bg,fontsize=\small]{c}
int NchooseK(int n, int k){
    if(k==0)return 1;
    if(n==k)return 1;
    else 
      return NchooseK(n-1, k-1) + 
            NchooseK(n-1, k);
}

	  \end{minted}
		\caption{A student's solution to part of an assignment in our dataset. 
    This function computes ${n \choose k}$ in C. }
    \label{fig:mossad_output:a}
	\end{subfigure}
        \hspace{1em}
    \begin{subfigure}[!t]{.45\textwidth}
		\begin{minted}[bgcolor=bg,fontsize=\small]{c}
int NchooseK(int n, int k){ 
  if(k==0) return 1; 
  int NchooseK(int n, int k);
  if(n==k)return 1;
  else
    return NchooseK(n-1, k-1) + 
          NchooseK(n-1, k);
  int i = 0;
}
	  \end{minted}
		\caption{A result of applying \systemfull{} to this code.\label{fig:mossad_output:b}}
	\end{subfigure}
    \begin{subfigure}{.47\textwidth}
        		\begin{minted}[bgcolor=bg,fontsize=\small]{text}
No matches were found in your submission.
	    \end{minted}
        \caption{Moss's output from comparing these programs.}
    \label{fig:mossad_output:c}
    \end{subfigure}
  \caption{\textbf{\systemfull{} in Action.} An example of
    \systemfull{}'s output, using the code in \ref{fig:mossad_output:a}
    as the input file and 10\% as the target Moss score. By injecting
    just two lines of semantics-perserving code, \systemfull{} successfully defeats Moss.}
\label{fig:mossad_output}
\end{figure}

\subsection{Score Checking}

The second check mutation performs is testing if the target Moss score
has been reached by querying Moss and scraping the web data from the
resulting URL. If the target has been achieved, then the current
generation will be the last and the temporary variant will be
outputted by \systemfull{}.

If the target Moss score check fails, the mutation module adds this
successful mutation to the pool to increase the odds of inclusion in
future variants of lines that have been proven to be successful. It
then returns to generation with the temporary variant as the new
variant.

\subsection{Program Equivalence}
\label{sec:program_equiv}

To determine program equivalence, the current prototype of
\systemfull{} simply directly compares the intermediate representation
of programs after compiling both with a high level of optimization. It
directly \emph{diff}s the resulting object code after compiling with
the \texttt{-O3} flag. \systemfull{} uses this method for a variety of
reasons: it is sound, fast, and is straightforward to implement.  This
approach to equivalence checking naturally limits the mutations to
those that would be compiled away, such as dead code and redundant
assignments.

This approach could, with additional engineering, be extended to
JIT-compiled interpreted languages by modifying the JIT-compiler to
emit generated object code. The key engineering obstacles to overcome
would be non-determinism in the JIT compilers themselves, such as when
they trigger optimization based on elapsed time rather than number of
instructions executed, or use other, pseudorandom sampling
approaches. Additionally, inlining across system library boundaries
would need to be disabled, as it could result in spurious matches.

Leveraging more advanced and general approaches for determining
program equivalence, such as parameterized program
equivalence~\cite{Kundu:2009:POC:1542476.1542513,
  Roychoudhury00verificationof} and program equivalence using
context-free grammars~\cite{BKRosen}, would open the door to a wider
range of mutations that could be applied. In particular, two variants
could produce different code while retaining semantic equivalence.
Another possible but unsound approach would be to generate a large
number of test cases via fuzzing, and using identical behavior on
those test cases as a proxy for equivalence.

\subsection{Entropy}

\label{sec:entropy}
To increase the likelihood of generating successful mutations,
\systemfull{} includes a feature that allows users to make additions
to the pool for selection, which we refer to as \emph{entropy}. Users
can add their own code to an external file that \systemfull{} will use
as an addition to the initial pool of the user input file. In
addition, if entropy collection is enabled, \systemfull{} will also
store all successful mutations to the entropy file during
mutation. This approach aims to increase the odds of successful
mutations in future uses of \systemfull{} for other assignments.

For our experiments (Section~\ref{sec:evaluation}), we manually
assembled a single entropy file based on our observations of common C
statements. These include common variables and basic mathematical
operations on them. For example, we included declarations and
assignments to variables
named \texttt{count}, \texttt{i}, \texttt{j}, \texttt{k}, \texttt{ret},
and \texttt{status}.  We found that these are extremely frequent in
public C codebases, such as Linux and Redis, making them generally
unsuspicious. The resulting entropy file contains approximately 30
lines of these statements.

\section{\systemname{} Evaluation}
\label{sec:evaluation}

Our evaluation focuses on the following research questions:

\begin{enumerate}

\item \textbf{RQ1 (Unsuspiciousness)}: Can \systemfull{} produce code variants that yield unsuspiciously low scores typical of non-plagiarized code?
\item \textbf{RQ2 (Readability)}: Does \systemfull{} produce code that is as no less readable (and thus no more suspicious) than legitimate programs?
\item \textbf{RQ3 (Effect of Input Size)}: How does the size of input programs affect \systemfull{}'s efficacy?
\item \textbf{RQ4 (Mass Plagiarism)}: Can \systemfull{} enable \emph{mass plagiarism} by producing numerous mutually dissimilar results from the same input programs?
\item \textbf{RQ5 (Performance):} Is the execution time required by \systemfull{} low enough to make it practical?
\item \textbf{RQ6 (Generality):} Is \systemfull{} equally effective at defeating other plagiarism detectors?
\end{enumerate}

\subsection{Methodology}

\subsubsection*{Dataset}
\label{sec:datasets}

To answer all of these research questions, we evaluate \systemfull{}
with course assignments provided by faculty from several universities;
all were previously stripped of personally identifiable
information. Our dataset consists of three homework projects from
undergraduate courses, all in C. Project 1 required students to
implement a bubblesort over an array of pointers to strings; solutions
range from 11--101 LOC (all figures computed by \texttt{sloccount},
which excludes comments and empty lines). Project 2 required students
to implement a program that converts parts of images (stored in a
format akin to PPM) from color to grayscale; solutions range from
19--53 LOC. Both Project 1 and 2 include 135 student
solutions. Project 3 required students to use recursion to implement a
function computing ${n \choose k}$.  This project includes 120 student
solutions, ranging from 14--116 LOC.

\subsubsection*{Suspiciousness Threshold}
\label{sec:threshold}

Evaluating \systemfull{}'s ability to produce unsuspiciously low
scores depends on the choice of a threshold of similarity that
constitutes suspiciousness. That is, programs that are below this
threshold of similarity will not be viewed as being likely to
constitute plagiarism, and thus will be less likely to be manually
inspected. Note that, for large classes, Moss does not report the
similarity of every file, instead reporting only the 250 most similar
pairs, in descending order of similarity.

For our evaluation, we empirically set a threshold that would result
in low false positive rates across our dataset. We examined the Moss
scores for each project in our dataset and found that the average
similarity score of assignments pairs (in the top 250 results) was
26.2\%, with a bootstrapped 95\% confidence interval of $[25.3\%,
27.2\%]$ (10,000 iterations); see Table~\ref{tab:scores} for complete
statistics. We manually inspected files in the high end of that range
and determined that these did not appear to constitute instances of
plagiarism. In addition, survey respondents indicated that the average
threshold they use for deeming Moss scores suspicious is
30\%. Informed by these two results, we conservatively set our
threshold for suspicion for experiments to 25\%, just below the bottom
of the confidence interval.

\subsection{RQ1: Unsuspiciousness}
\label{sec:mossad_effectiveness}

\begin{figure*}[tb]
	\centering
        \begin{subfigure}[t]{0.34\linewidth}
          \includegraphics[width=0.99\textwidth]{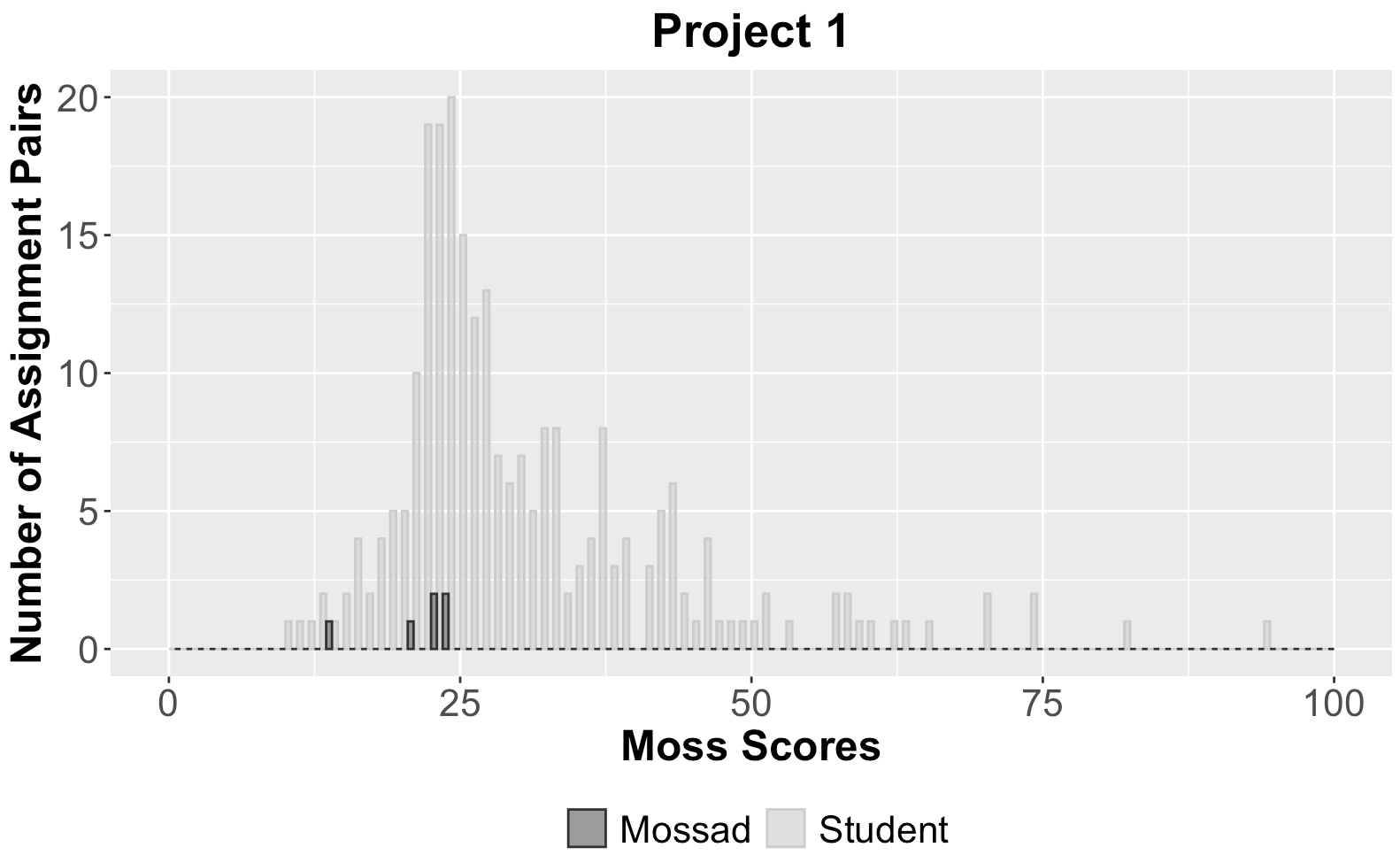}
          \caption{}
          \label{fig:lab4_source}
        \end{subfigure}\hfill%
        \begin{subfigure}[t]{0.33\linewidth}
          \includegraphics[width=0.99\textwidth]{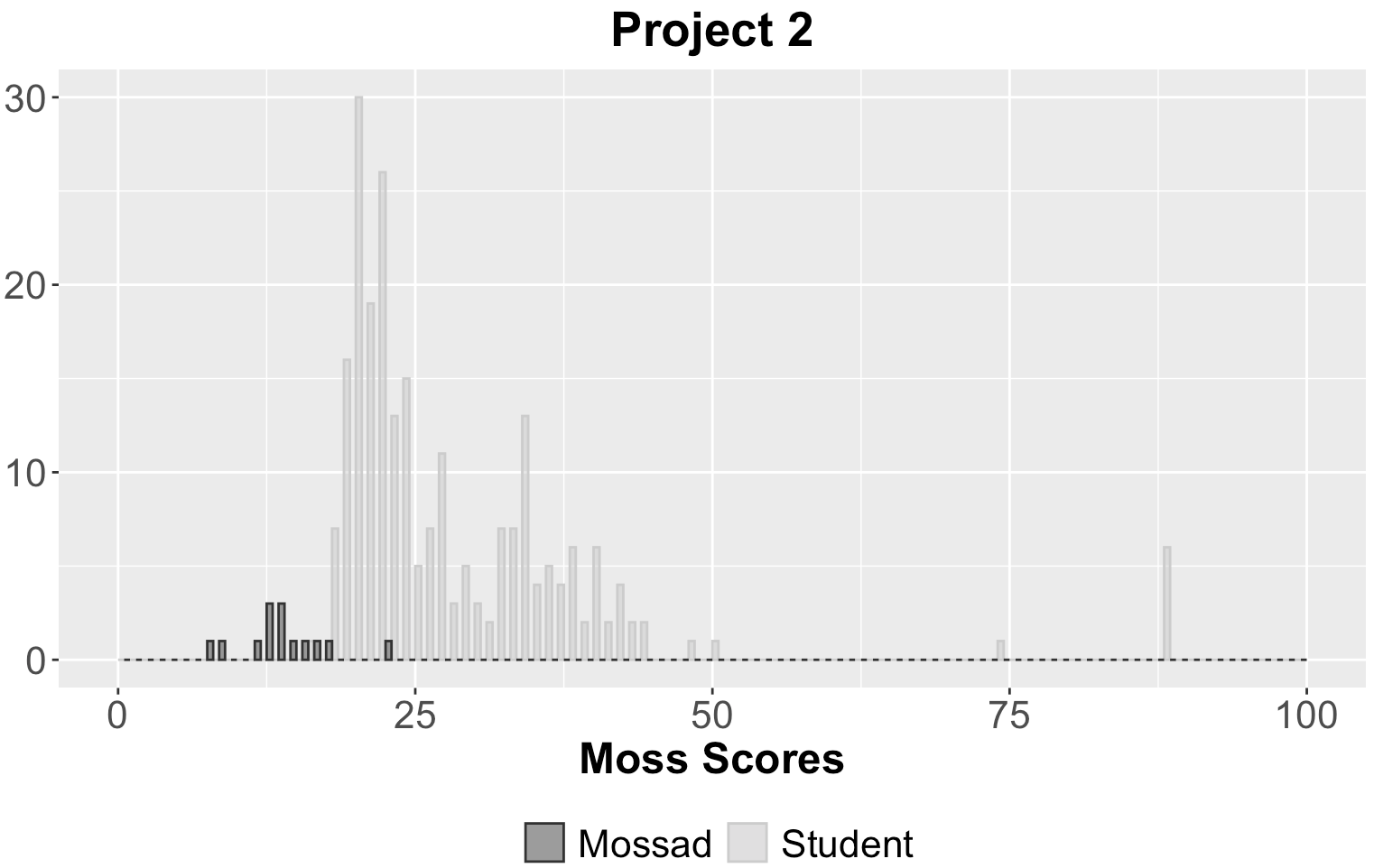}
          \caption{}
          \label{fig:lab7_source}
        \end{subfigure}\hfill%
        \begin{subfigure}[t]{0.33\linewidth}
          \centering\includegraphics[width=0.99\textwidth]{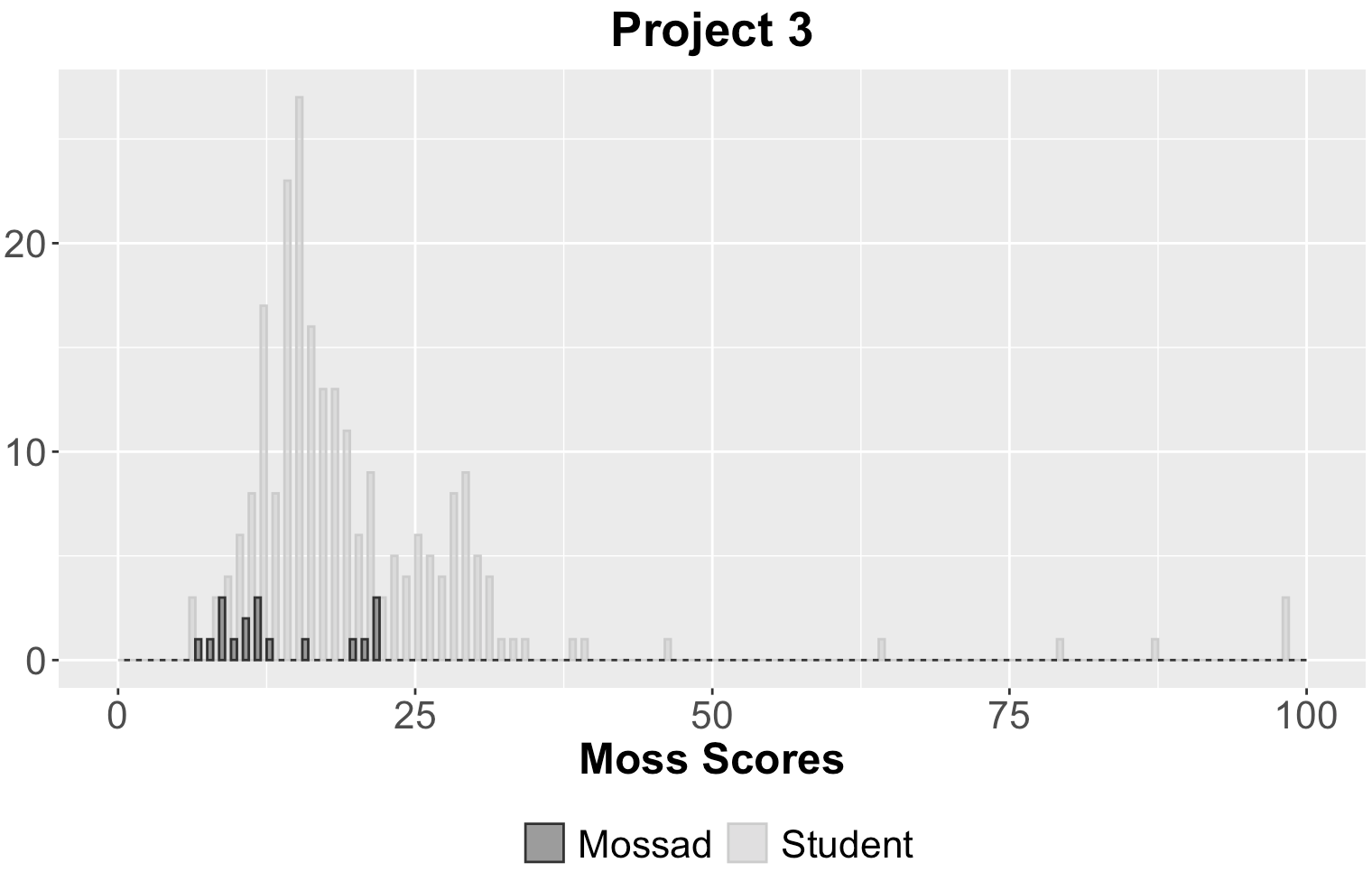}
          \caption{}
          \label{fig:lab8_source}
        \end{subfigure}
  \caption{\textbf{Across a suite of actual student assignments, \systemfull{} consistently produces variants that result in unsuspicious Moss scores.} Light gray indicates legitimate assignments, while dark gray indicates \systemfull{}-produced variants. ($\S\ref{sec:mossad_effectiveness_scores}$)}
  \label{fig:source_all}
\end{figure*}

The most important metric of performance for a system aimed at
defeating plagiarism is its effectiveness at producing unsuspicious
variants; that is, the similarity of transformed programs to the
original version is reported as below the suspiciousness threshold we
derive above (25\%). To measure \systemfull{}'s effectiveness, we
apply \systemfull{} to files from three C data sets and used Moss to
compare the generated variants against the entire body of student
code.

\subsubsection{Moss Scores for \systemfull{}-Produced Code}
\label{sec:mossad_effectiveness_scores}

We perform the same set of experiments across the projects in our dataset.
For each, we randomly select five assignments to be used as inputs to
\systemfull{}.  We run \systemfull{} with a target Moss score of
25\% together with the entropy file described in Section~\ref{sec:entropy}. After
generating these \systemfull{} variants, we add these five additional
files, and then ran Moss across the entire corpus.

\systemfull{} performed similarly across all of the projects;
Figure~\ref{fig:source_all} presents these results. As
previously mentioned, Moss returns a list of the highest scoring pairs
of assignments; our graphs present the scores of these pairs. For each
pair consisting of least one \systemfull{} variant, the corresponding
bins are shaded in dark gray, while for each pair that consists solely
of authentic student code, the corresponding bins are shaded in light
gray.

\begin{table}[!t]
\begin{tabular}{lrrrr}
                   & \multicolumn{4}{l}{\textbf{Moss similarity scores}} \\
\textbf{Code type} & \textbf{\emph{Min}} & \textbf{\emph{Max}} & \textbf{\emph{Avg}} & \textbf{\emph{95\% CI}} \\
\toprule
\multicolumn{5}{c}{\textbf{\emph{Individual plagiarism}}} \\
\emph{\systemfull{}} & 7\% & \textbf{24\%} & \textbf{15.4\%} & [\textbf{13.6\%}, \textbf{17.2\%}] \\
\emph{\systemclone{}} ($\S\ref{sec:mossad_det}$) & \textbf{6\%} & 31\% & 16.7\% & [13.8\%, 19.7\%] \\
\emph{\systemshatter{}} ($\S\ref{sec:mossad_nondet}$)  & \textbf{6\%} & 33\% & 19.6\% & [14.5\%, 24.7\%] \\
\hline
\multicolumn{5}{c}{\textbf{\emph{Mass plagiarism}}} \\
\emph{\systemfull{}} ($\S\ref{sec:mossad_mass_plagiarism}$)  & \textbf{5\%} & \textbf{27\%} & \textbf{12.1\%} & [\textbf{11.5\%}, \textbf{12.7\%}] \\
\emph{\systemshatter{}} ($\S\ref{sec:mossad_nondet}$)  & 15\% & 46\% & 24.2\% & [23.4\%, 25.0\%] \\
\hline
\emph{Legitimate student code} & 6\% & 98\% & 26.2\% & [25.3\%, 27.2\%] \\
\end{tabular}
\vspace{1em}
\caption{\textbf{Summary of effectiveness results for \systemfull{} ($\S\ref{sec:mossad}$) and ``ablated'' variants ($\S\ref{sec:ablation}$)}. Lowest values are shown in \textbf{boldface}. All of the variants produce code that is usually below the average similarity of non-plagiarized student code. \systemfull{} always yields the lowest similarity scores. \emph{Individual plagiarism} denotes an attempt to plagiarize from exactly one source program; \emph{mass plagiarism} denotes an attempt to produce multiple variants from a single source program. Note that \systemclone{} is ineffective for mass plagiarism, since its transformations are deterministic, as all generated versions would be identical ($\S\ref{sec:ablation}$).\label{tab:scores}}
\end{table}

Table~\ref{tab:scores} summarizes the average similarity scores and
their bootstrapped 95\% confidence intervals. For every project,
the \systemfull{} variants achieved similarity scores lower than the
average Moss score for the 250 highest scoring pairs of authentic
programs. \systemfull{} yielded variants with average similarity
scores of 15.4\%, while the average similarity of legitimate code was
26.2\%. As with the legitimate student code, we used the bootstrap to
compute a 95\% confidence interval around the means; the confidence
interval for \systemfull{} variants was $[13.6\%, 17.2\%]$.

\definecolor{shadecolor}{rgb}{0.85,0.85,0.85}
\begin{shaded}
\textbf{RQ1 (Unsuspiciousness)}: \emph{\systemfull{} generates variants that are unsuspicious: on average, Moss computes their similarity to the original versions to be just 15\%, well below the average similarity of authentic programs.}
\end{shaded}

\subsection{RQ2: Readability of \systemfull{}-Generated Code}
\label{sec:mossad_grading}
\label{sec:neena_experiment}

While generating unsuspicious variants is important, it is not enough
to avoid the risk of detection in case of spot checks or routine
manual inspection. We examine \systemfull{}'s effectiveness at
producing variants that are just as readable as authentic student code
via an experiment with teaching assistants in-training. For the purposes
of this evaluation, the term ``readable'' is used roughly to mean that the
code looks like legitimate student code to graders.

For the user study, the participants ($N$=30) were Computer Science PhD students
enrolled in a mandatory teacher training course, required before they can serve
as Teaching Assistants at their home university. Participants were in
the first to fourth years of their PhD.

To simulate the effect of manually grading \systemfull{} code in the
real world, participants were asked to grade a series of student
assignments consisting of a mix of authentic student
code, \systemfull{}-generated code, and obfuscated code. Each participant was
given assignments in random order. All assignments were anonymized. We
deliberately did not provide Moss similarity scores; because \systemfull{}
generates extremely low scores, we did not want to risk biasing
students against detecting plagiarism for these cases.

The participants were provided the prompt for the assignment along with a
rubric that contained three core elements that needed to be scored on
a four-point scale (1-4, higher is
better): \emph{Accuracy}, \emph{Program Design}, and \emph{Code
Readability}. This grading rubric was derived from several publicly
available grading rubrics for Computer Science courses. Each participant
was given 20 minutes to grade 10 assignments in order to mimic the
average time a grader actually spends manually inspecting code, as
reported by survey respondents.

The rubric provided guidance for each score for each task and asked
the participants to write a few sentences on their overall impression of
the code, including anything they believed the instructor should
know. The participants were given time to familiarize themselves with the
rubric before they were given assignments to grade. For
the Accuracy component, participants were given a
simulated autograder score for each assignment. For Program
Design, participants were asked to judge the structures used and the
overall design of algorithms. Finally, for Readability, participants were
instructed to examine the documentation and indentation of the code,
as well as if the identifiers used were appropriate for the task and
if the code was overall easy to read.

As in all of the experiments in this evaluation, the \systemname{}
code presented to participants contained only additional trivial
statements and declarations, all of which optimize away during
compilation (i.e., they are semantics-preserving).

\textbf{Results:} Using Hedges' G on the two populations of authentic student code
scores and \systemfull{} code scores, we find the effect size for both
Program Design and Code Readability to be low (0.15 and 0.11,
respectively). That is, the effect of using \systemname{} on programs
results in little impact on their assessed program design or
readability. Participants were asked to provide a few sentences in
feedback for each assignment and to relay notes to the instructor if
they felt the need; none reported anything out of the ordinary, let
alone suspicious code or plagiarism.

However, using Hedges' G with the populations of authentic student
code scores and obfuscated code scores, we discovered very different,
extreme results. In this case, the effect of obfuscation was quite
large on both Program Design (1.7) and Code Readability (1.1),
empirically substantiating our expectation that obfuscated code looks
markedly different from unobfuscated code, as anticipated in
Section~\ref{sec:obfuscator_eval}.

\begin{shaded}
\textbf{RQ2 (Readability)}: \emph{\systemfull{} generates variants that are essentially indistinguishable from authentic programs in terms of readability.}
\end{shaded}

\subsection{RQ3: Effect of Input Size on \systemfull{}}
\label{sec:mossad_generalizable}

To assess the impact of input size (that is, the length of input
programs) on \systemfull{}, we evaluate \systemfull{}'s effectiveness
at generating variants from input files of varying
sizes, with and without an entropy file.

Throughout the experiments, input file size was observed to have an
effect on both Moss scores and \systemfull{} outputs. With Moss, for
small enough files, it appears that the window size for hashing is
simply too large, since very minimal disruptions are needed to reach a
0\% Moss score. It is intuitive that smaller files need fewer
disruptions; however, the number of disruptions does not follow a
linear scale for achieving low Moss scores. We observe that the number
of disruptions for small files to achieve low scores is not
proportional to the number of disruptions for large files.

The length of program inputs has a variety of effects
on \systemfull{}. We first consider these assuming \systemfull{} does
not use its entropy file feature. For small files, \systemfull{}
performs poorly without an entropy file. Although smaller files
need a proportionally lower amount of insertions to disrupt the
amount of windows, the number of successful mutation possibilities
are extremely low. Small files generally have fewer
semantics-preserving lines of code, so it limits the possible lines
that can be re-inserted by \systemfull{}. Smaller files also have
limited locations for insertions such that the insertions do not
affect the output of the program. On the other hand, for larger files,
there are more mutation possibilities: more possible statements that
can be inserted, and more locations where such statements can be
inserted.

We empirically evaluate the length of input files necessary
for \systemfull{} to run properly without the need for an entropy
file. For this experiment, we sorted the programs from our data sets
by length (LOC) and ran \systemfull{} on each assignment in descending
order until \systemfull{} consistently timed out: we define these terms below.

Since \systemfull{} is nondeterministic, we used each assignment as an
input four times. We set our timeout threshold at five
minutes. If \systemfull{} could not find a single successful mutation
within that time, or if the variant ever reaches a length of
2.5$\times$ the length of the input file without achieving a Moss
score that meets the threshold, it would then timeout.
We deem ``consistent timeouts'' as
those that timeout for half or more of the iterations, for three
unique assignments in a row. Using these methods, we found that, on
average, \systemfull{} is not effective on inputs of 35 LOC or less
without an entropy file.

When \systemfull{} is run with an entropy file, the system produces
target-achieving outputs regardless of input size.  An entropy file
increases the probability of generating successful
mutations, by definition, which is independent of input size. In
fact, the experiment to evaluate this is exactly the experiment
shown previously in Figure~\ref{fig:source_all}, in which
\systemfull{} consistently produces variants that achieve Moss scores
below 25\%.

\begin{shaded}
\textbf{RQ3 (Effect of Input Size)}: \emph{\systemfull{} is effective on programs above 35 LOC; for shorter programs, it is effective when incorporating an entropy file consisting of generic lines of program source code.}
\end{shaded}

\subsection{RQ4: Mass Plagiarism with \systemfull{}}
\label{sec:mossad_mass_plagiarism}

\begin{figure}[!t]
    \centering
    \includegraphics[width=80mm]{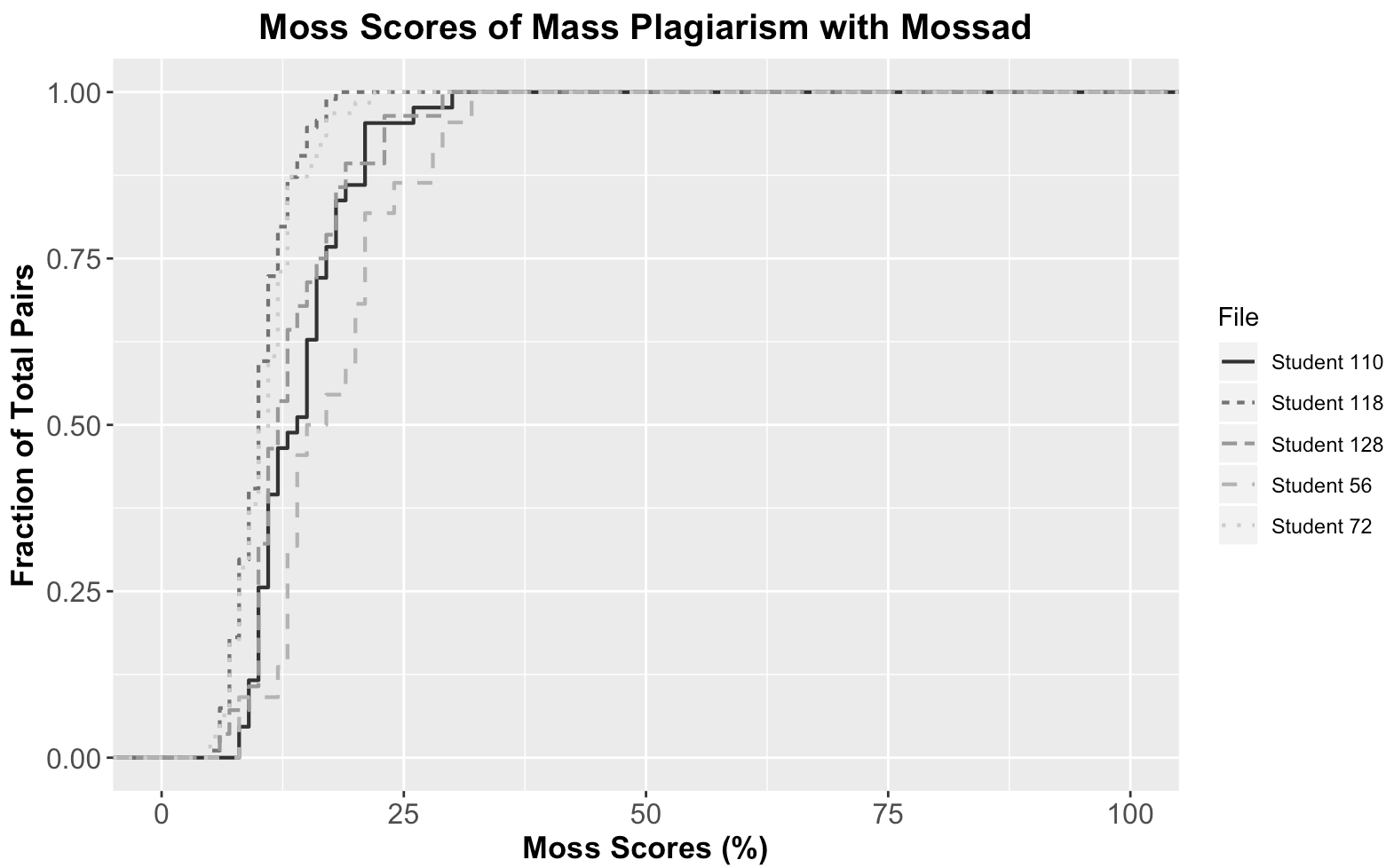}
    \caption{\textbf{Mass plagiarism with \systemfull{} yields unsuspicious Moss scores.}
    Because \systemfull{} is non-deterministic, it can generate multiple variants from the same input source code. Using 5 different base files as input to \systemfull{} and creating 30 variants per base file, we find that the highest Moss score of any pair of files is 27\%, not significantly larger than the 25\% suspiciousness threshold.}
    \label{fig:mass_full}
\end{figure}

To demonstrate the effects of mass plagiarism with \systemfull{}, our
third experiment consisted of generating
multiple \systemfull{}-disguised files from the same input file and
using Moss to assess the similarity scores between said variants. Using
five randomly-selected student assignments from our data sets as
input, we generated 30 \systemfull{} variants per base
file. Since \systemfull{} is nondeterministic, this process created 30
mutually-dissimilar files per base file. Separately, each set
of 30 variants was assigned similarity scores by Moss; these scores
are shown in Figure~\ref{fig:mass_full}. Recall that Moss only shows
the top 250 highest-scoring pairs, therefore, for each base file
there are 250 corresponding data points (1250 in total).

Since \systemfull{} is completely
nondeterministic, the odds of an insertion hit between files is
supremely low, resulting in overall low Moss scores. As shown in
Figure~\ref{fig:mass_full}, 90\% of the top 250 pairs score below 26\%
similarity for each unique base file, which is only 1\% higher than the
threshold of suspicion. The highest Moss score was 27\% and the average
score was 12.1\%. These results are summarized in Table~\ref{tab:scores}.

\begin{shaded}
\textbf{RQ4 (Mass Plagiarism)}: \emph{\systemfull{} is able to generate large numbers of variants from a base file that Moss scores as non-suspicious, enabling \emph{mass plagiarism}.}
\end{shaded}

\subsection{RQ5: \systemfull{}: Performance}
\label{sec:mossad_performance}

The performance of \systemfull{} primarily depends on the latency and
threshold of Moss, which is a function of the number of jobs at any
particular moment. In this section, we examine the run time
performance of \systemfull{} for each of our three data sets. For each
data set, three randomly selected student assignments are used as
inputs to \systemfull{}, along with the default Moss score of 25\% we
established in Section~\ref{sec:threshold}, and each input is
evaluated with \systemfull{} ten times.

\begin{figure}[!t]
    \centering \includegraphics[width=80mm]{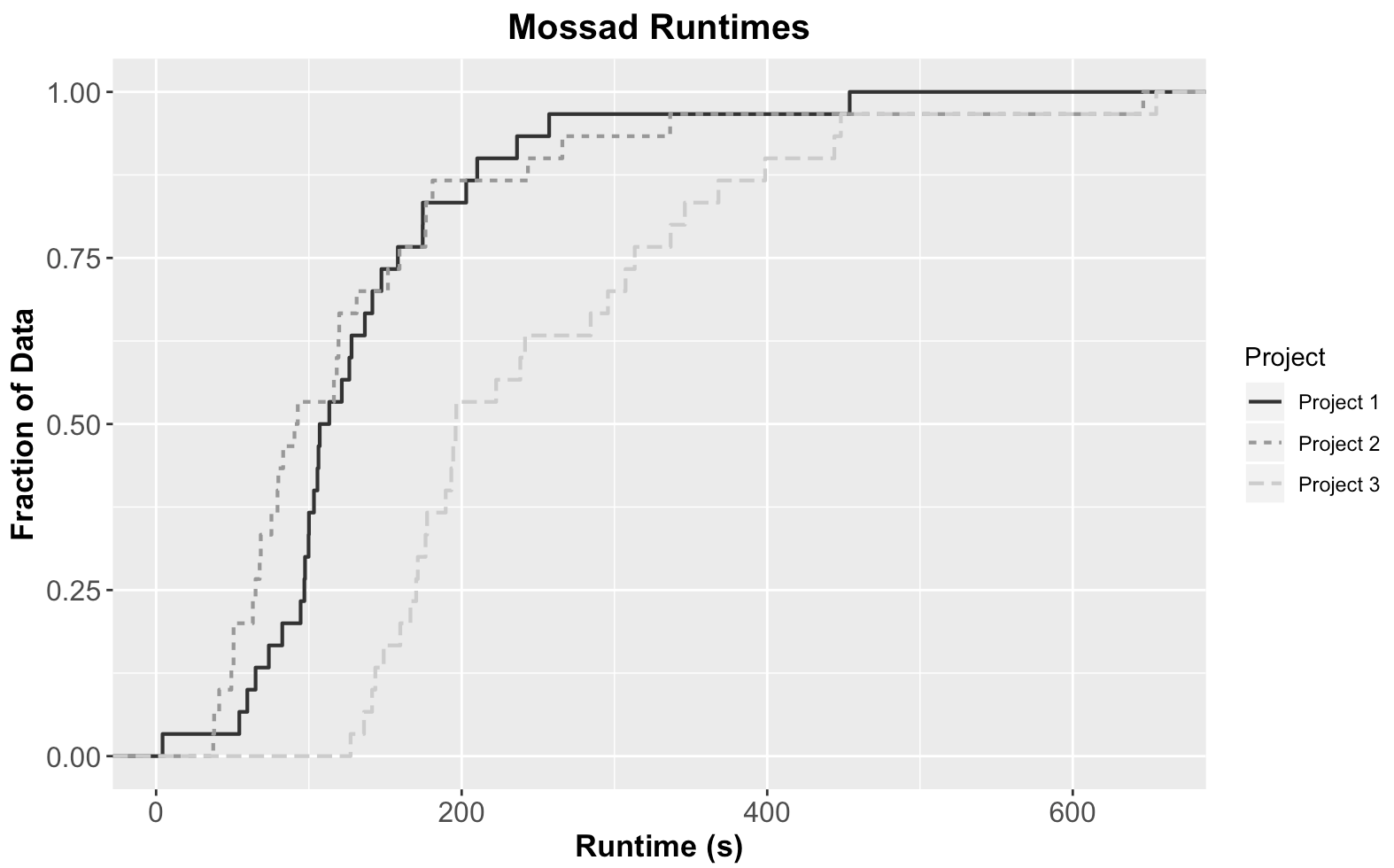} \caption{ \textbf{\systemfull{}
    consistently takes less than 5 minutes to create a
    variant.} 95\% of Project 3 iterations complete within 7
      minutes, the longest of all three data sets.}
    \label{fig:cdfs}
\end{figure}

Figure~\ref{fig:cdfs} plots the distribution of execution times for
all 30 iterations for each project. As shown in the graph, Project 3
inputs produced the longest run times with 95\% completing within 7.5
minutes. Inputs from Projects 1 and 2 resulted in very similar run
times for almost all of their iterations, with 95\% of the iterations
completing within approximately 4 minutes for Project 1 and 5.5
minutes for Project 2. These run times are nondeterministic and can
change depending on how many jobs are running simultaneously on
Moss. That said, these times are in general far less than the time
required to actually do the homework assignments.

\begin{shaded}
\textbf{RQ5 (Performance)}: \emph{\systemfull{} is able to generate variants with low similarity scores in under 10 minutes, far less than the amount of time generally allotted to complete assignments.}
\end{shaded}

\subsection{RQ6: \systemfull{} vs. Other Plagiarism Detectors}
\label{sec:othertools}

To address the question of \systemfull{}'s effectiveness against other
software plagiarism detection tools, aimed to collect all of the systems discussed in Section~\ref{sec:relatedwork}. Unfortunately, this effort was
largely unsuccessful, as almost none of these systems are available for use. We reached out to the authors of the cited papers, and were unsuccessful in obtaining any of their systems, with three exceptions: 
JPlag~\cite{prechelt2002finding},
Sherlock~\cite{joy1999plagiarism},
and Fett~\cite{syntaxbased}. We focus our evaluation on JPlag, since it
was the second-most popular tool cited by our survey respondents. Only one
respondent reported using Sherlock, and none reported using Fett.

\subsubsection{JPlag}
\label{sec:jplag}
Like Moss, JPlag is a system for detecting similarity of software;
both systems use tokenization, but instead of hashing, JPlag uses an
approach based on string tiling to identify matches. From our survey
results, we found that JPlag was the second most common software
plagiarism tool, though it was a distant second to Moss. 

To examine \systemfull{}'s effectiveness on JPlag, we performed the
same experiment as outlined in Section~\ref{sec:mossad_effectiveness},
replacing Moss with JPlag as the similarity detector. We refer
to \systemfull{} augmented with JPlag as
\systemjplag{}.

Unlike Moss, which returns the top 250 highest scoring pairs of
results, JPlag returns \emph{all} possible pairwise combinations of
input assignments. As a result, the \systemjplag{} experiments result
in many more observable scores. We repeat all experiments with JPlag
with the same threshold used in the previous experiments with Moss
(25\%).

\begin{figure*}[!t]
	\centering
        \begin{subfigure}[t]{0.34\linewidth}
          \includegraphics[width=0.99\textwidth]{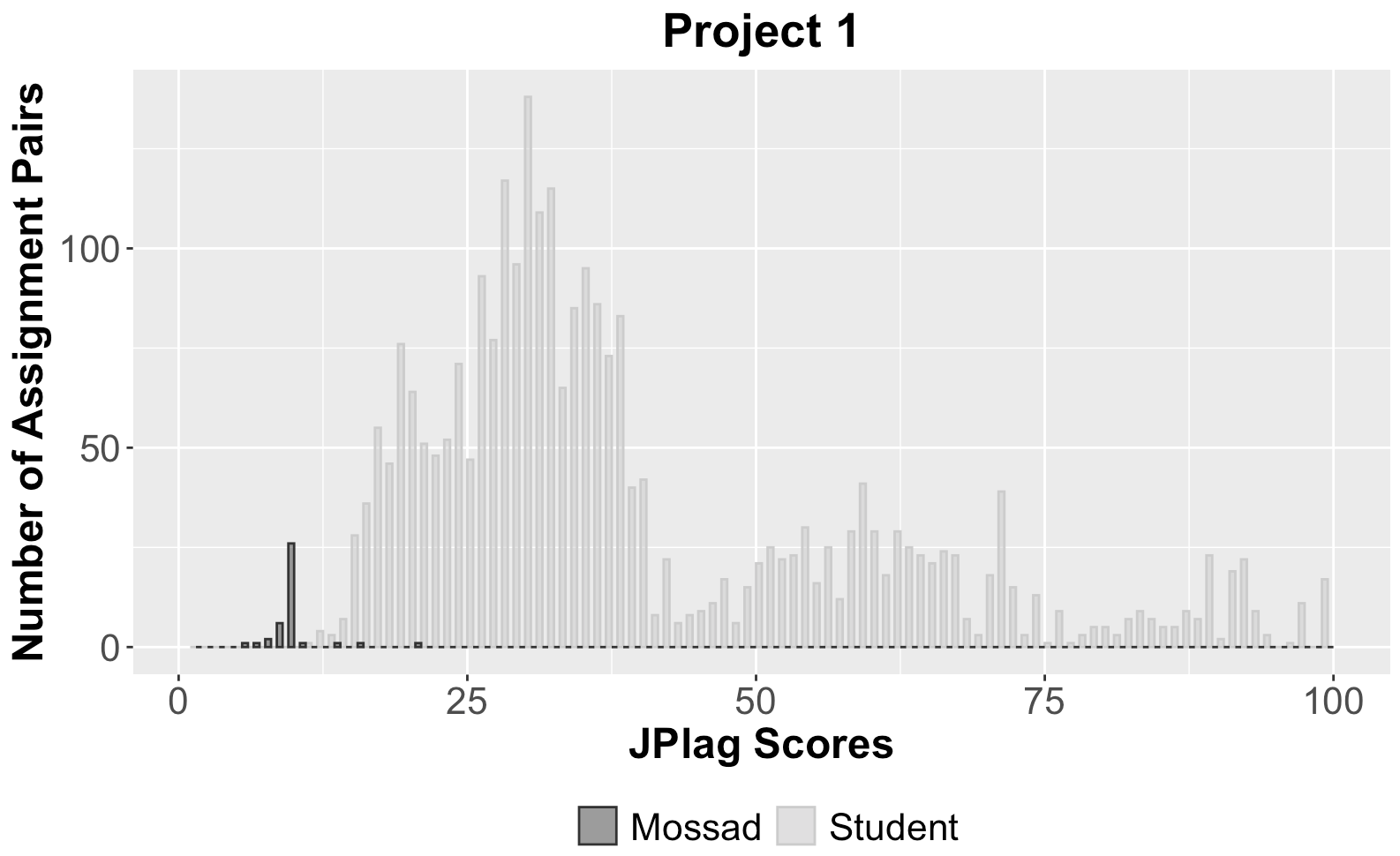}
          \caption{\textbf{Project 1.}}
          \label{fig:lab4_jplag}
        \end{subfigure}\hfill%
        \begin{subfigure}[t]{0.33\linewidth}
          \includegraphics[width=0.99\textwidth]{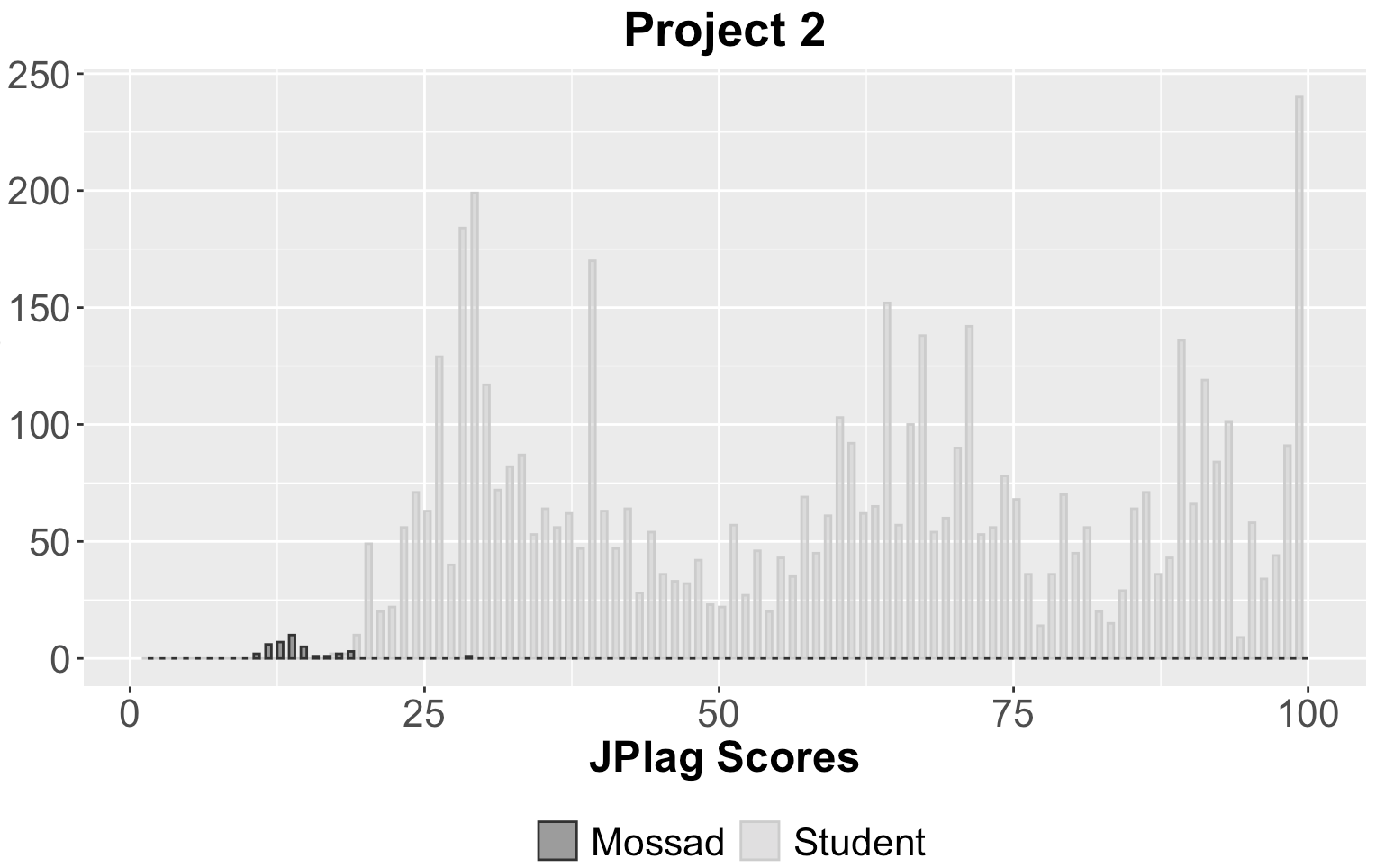}
          \caption{\textbf{Project 2.}}
          \label{fig:lab7_jplag}
        \end{subfigure}\hfill%
        \begin{subfigure}[t]{0.33\linewidth}
          \centering\includegraphics[width=0.99\textwidth]{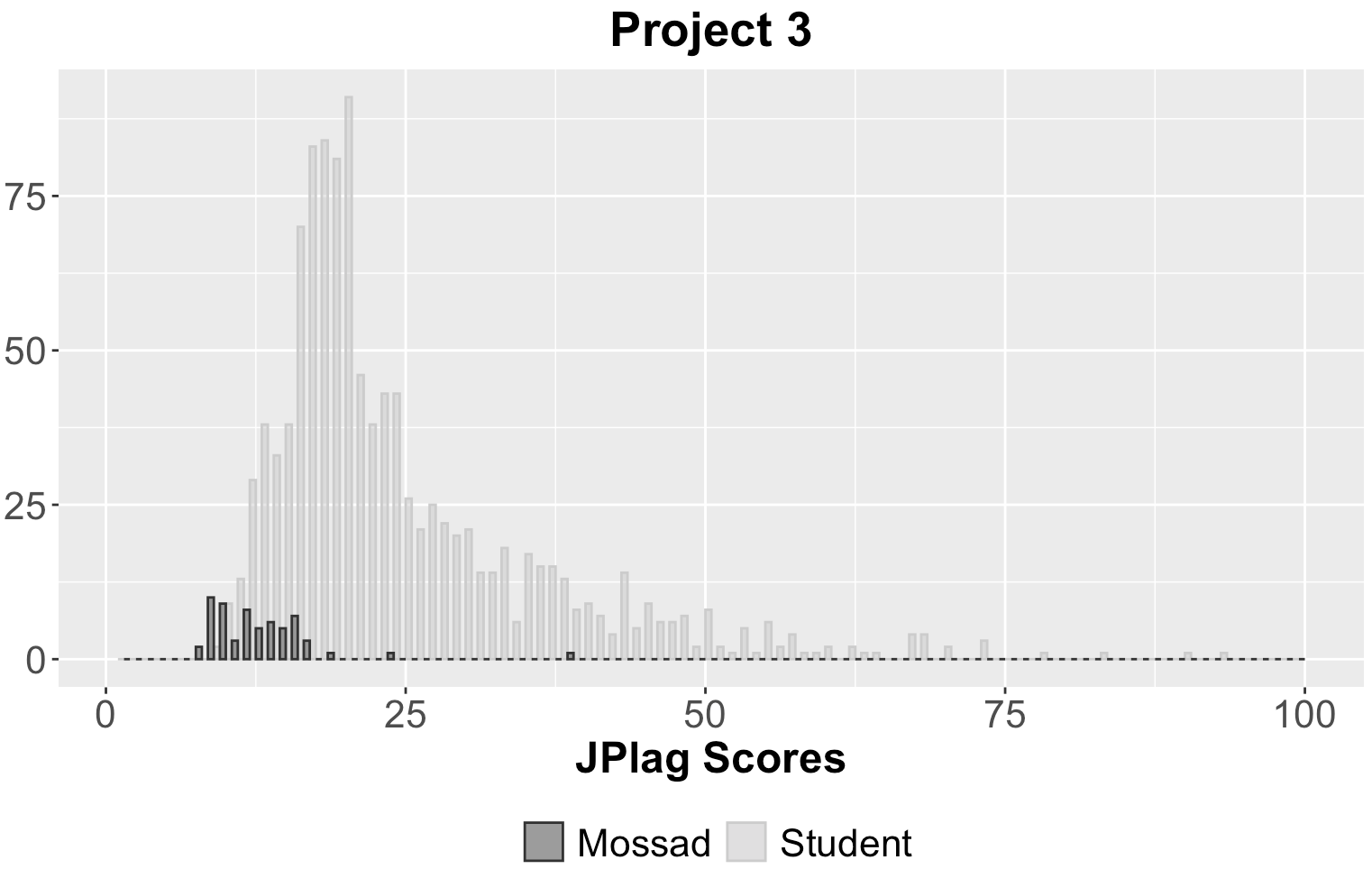}
          \caption{\textbf{Project 3.}}
          \label{fig:lab8_jplag}
        \end{subfigure}
  \caption{\textbf{\systemfull{} effectiveness versus JPlag~\cite{prechelt2002finding}, another widely used software plagiarism detector.} \systemfull{}-produced variants produce even lower similarity scores with JPlag than with Moss, with one outlier at 38\% in Project 3. For each dataset, we omit pairs scoring 0\%, which constitute the vast majority of results.}
  \label{fig:jplag_all}
\end{figure*}

The results from executing JPlag on all three projects are shown in
Figure~\ref{fig:jplag_all}. The vast majority of the pairs ($> 8000$)
scored 0\%; we omit these results from these graphs for clarity. As
before, the scores for pairs of assignments with at least
one \systemjplag{} file are shown in dark gray, and the scores for
authentic student assignment pairs are shown in light gray.

Table~\ref{tab:jplag_scores} summarizes our results. For all three
projects, JPlag resulted in average similarity scores that were
substantially higher than Moss. Just as
with \systemfull{}, \systemjplag{} was successful at producing results
whose similarity scores are noticeably lower than the bulk of the
distribution of legitimate programs, even with the 0\% matches
excluded.

\systemjplag{} produces variants that are below the 25\% suspiciousness threshold across the suite,
with one exception: a case where \systemjplag{} produced a variant
that scored 38\% with JPlag. We note that this value crosses a
suspiciousness threshold that we derived for Moss. However, given that
JPlag's scores are much higher (the average similarity of the top 250
matches was 76.9\%, this threshold is probably too low for JPlag. That
is, we expect users of JPlag would stop inspecting at far higher
thresholds. Even including the 0\% matches, the average JPlag score
assigned to legitimate student code is 28\%, whereas the average score
assigned to \systemjplag{} scores is 1\%.

\begin{table}[!t]
\begin{tabular}{lrrrr}
\textbf{Code type} & \textbf{\emph{Min}} & \textbf{\emph{Max}} & \textbf{\emph{Avg}} & \textbf{\emph{95\% CI}} \\
\toprule
\multicolumn{5}{c}{\textbf{JPlag similarity scores}} \\
\multicolumn{5}{c}{\textbf{\emph{all matches}}} \\
\emph{\systemjplag{}} & \textbf{0\%} & \textbf{38\%} & \textbf{1.1\%} & [\textbf{0.9\%}, \textbf{1.3\%}] \\
\emph{Legitimate Student Code} & 0\% & 100\% & 28.2\% & [27.7\%, 28.7\%] \\
\multicolumn{5}{c}{\textbf{\emph{top 250 matches}}} \\
\emph{Legitimate Student Code} & 30\% & 100\% & 76.9\% & [74.7\%, 78.3\%] \\
\hline
\multicolumn{5}{c}{\textbf{Sherlock similarity scores}} \\
\emph{\systemfull{}} & \textbf{1}\% & \textbf{61}\% & \textbf{6.2}\% & [\textbf{5.7\%}, \textbf{6.8\%}] \\
\emph{Legitimate Student Code} & 2\% & 100\% & 24.1\% & [23.8\%, 24.4\%] \\
\hline
\multicolumn{5}{c}{\textbf{Fett similarity scores}} \\
\emph{\systemfull{}} & \textbf{0.6}\% & \textbf{37.6}\% & \textbf{5.9}\% & [\textbf{4.7\%}, \textbf{7.4\%}] \\
\emph{Legitimate Student Code} & 2.3\% & 100\% & 45\% & [38.6\%, 50.4\%] \\
\end{tabular}
\vspace{1em}
\caption{\textbf{Summary of effectiveness results against three other plagiarism detectors: JPlag (with the complete JPlag output and only the top 250 pairs, for ease of comparison with Moss), Sherlock, and Fett.}\label{tab:jplag_scores}}
\end{table}

For Project 1, the average JPlag scores hover around 30\%, and
the \systemjplag{} scores center around a mean of 12\%
(Figure~\ref{fig:lab4_jplag}). One interesting result of using JPlag on
this project is a higher percentage of 100\% matches within the
dataset without \systemjplag{} variants. In fact, 174 files across
the 3 datasets received 100\% similarity scores, and 536 scored
above 97\%. We randomly chose a subset (5) of the 100\%-scoring
pairs to spot check in order to assess the true positive rate.
From these pairs, we found that approximately half exhibit high
similarity in terms of algorithm and text, which we believe could
be reasonably flagged as suspiciously similar. The other half of the
files, although similar, had different algorithms and were 
not clearly plagiarized. 

For Project 2, the JPlag results do not exhibit a clear drop off
between the average student assignment similarity score and the
outliers that we observed in the \systemfull{} experiments using Moss
(Figure~\ref{fig:lab7_jplag}). Instead, there are three clusters of
scores at 30\%, 65\%, and 90\%. The mean \systemjplag{} score is 10\%.

Finally, for Project 3, the JPlag scores follow a heavy-tailed
distribution (Figure~\ref{fig:lab8_jplag}). Here, the \systemjplag{}
scores are not significantly lower than the mean score of the
legitimate solutions, approximately 20\%.

\subsubsection{Sherlock}
\label{sec:sherlock}

Sherlock~\cite{joy1999plagiarism} is the third-most popular similarity detection
system that is publicly available; it was cited
by one survey respondent.  It is a general-purpose
academic plagiarism detector with support for both source code as well
as natural language. For source code, it translates each source file
into three \textit{modes}: the original document, normalized
(whitespace and comments are removed), and tokenized, where the code
is parsed into tokens according to their basic purpose (similar to
Moss's normalization pass).

For each mode, Sherlock compares every pair of assignments and 
calculates a similarity score based on the total number of lines. However,
Sherlock only displays to the user the sum of the total mode scores.

Using \systemfull{}-generated variants for each of our three datasets,
we compared the entire corpus augmented with the \systemfull{} code
using Sherlock. Since Sherlock reports a sum of the scores for each
mode, we normalize its output simply by dividing by 3 (the system
recommended leaving the mode settings untouched when using
C).

\systemfull{} is generally as effective against Sherlock
as against the other plagiarism detectors we examine. While in
one case, a \systemfull{} assignment receives a score of 61\%, the
average for \systemfull{}-generated code is 6.2\%, well below the
average score for legitimate student code (24\%); the 95\% confidence
interval for \systemfull{} code is [5.7\%,
6.8\%]. Table~\ref{tab:jplag_scores} provides a summary of these
statistics.

\subsubsection{Fett}
\label{sec:fett}
Fett~\cite{syntaxbased} is a recent system aimed at addressing the drawbacks
of Moss, JPlag, and other related plagiarism detection tools. Although none
of our survey respondents reported using the tool, Fett is the only additional
publicly-available plagiarism detection tool the authors could find. In fact,
one of the major pillars of Fett is that it is open-source and easily accessible, along with being language-agnostic and highly accurate~\cite{syntaxbased}.

Fett linearizes the resulting parse tree generated by an ANTLR
parser via postorder traversal~\cite{10.1145/3159450.3159490}. It then sorts the
functions of the program by size, prunes the sequences, groups parse
nodes into equivalence classes, and assigns weights to remaining nodes
based on their class. Then, Fett uses the Smith-Waterman algorithm for
sequence alignment scoring, penalizing gaps. Fett generates a matrix of
similarity scores in CSV format.

Performing the same experiment described in Section~\ref{sec:sherlock}, using
the same C dataset and \systemfull{}-generated variants,
we found that \systemfull{} is generally as effective against Fett
as the other systems we evaluate.

The largest score assigned to a pair of files, where at least one file in the 
pair is a \systemname{} variant, is 37\%. The
average for \systemfull{}-generated code is 5.9\%, well below the
average score for legitimate student code (45\%); the 95\% confidence
interval for \systemfull{} code is [4.7\%, 7.4\%].
These statistics are summarized in Table~\ref{tab:jplag_scores}.

Although \systemname{} in its current form is able to completely
undermine Fett for C datasets, \systemname{} would need to be altered
slightly in order to have the same impact with Java code. For Java,
Fett removes all expressions~\cite{syntaxbased}, which eliminates the
kind of code that \systemname{} inserts by default. A straightforward
workaround is to replace the expressions that \systemname{} inserts
with alternative pieces of code that would not be eliminated,
including empty if-statements or calls to functions with no side
effects.

\begin{shaded}
\textbf{RQ6 (Generality)}: \emph{\systemfull{} is effective against other plagiarism detectors (JPlag, Sherlock, and Fett), consistently achieving low similarity scores.}
\end{shaded}

\subsection{Ablation Study}
\label{sec:ablation}

To isolate the effects of \systemfull{}'s algorithmic approaches, we
conduct an ablation study. We constructed two variants
of \systemfull{} that we evaluate here. Note that,
unlike \systemfull{}, neither of these rely on Moss itself. Both of
these approaches also directly perform hash disruption by targeting
fingerprint windows, while \systemfull{} does so implicitly.

\begin{figure*}[!t]
	\centering
        \begin{subfigure}[t]{0.34\linewidth}
          \includegraphics[width=0.99\textwidth]{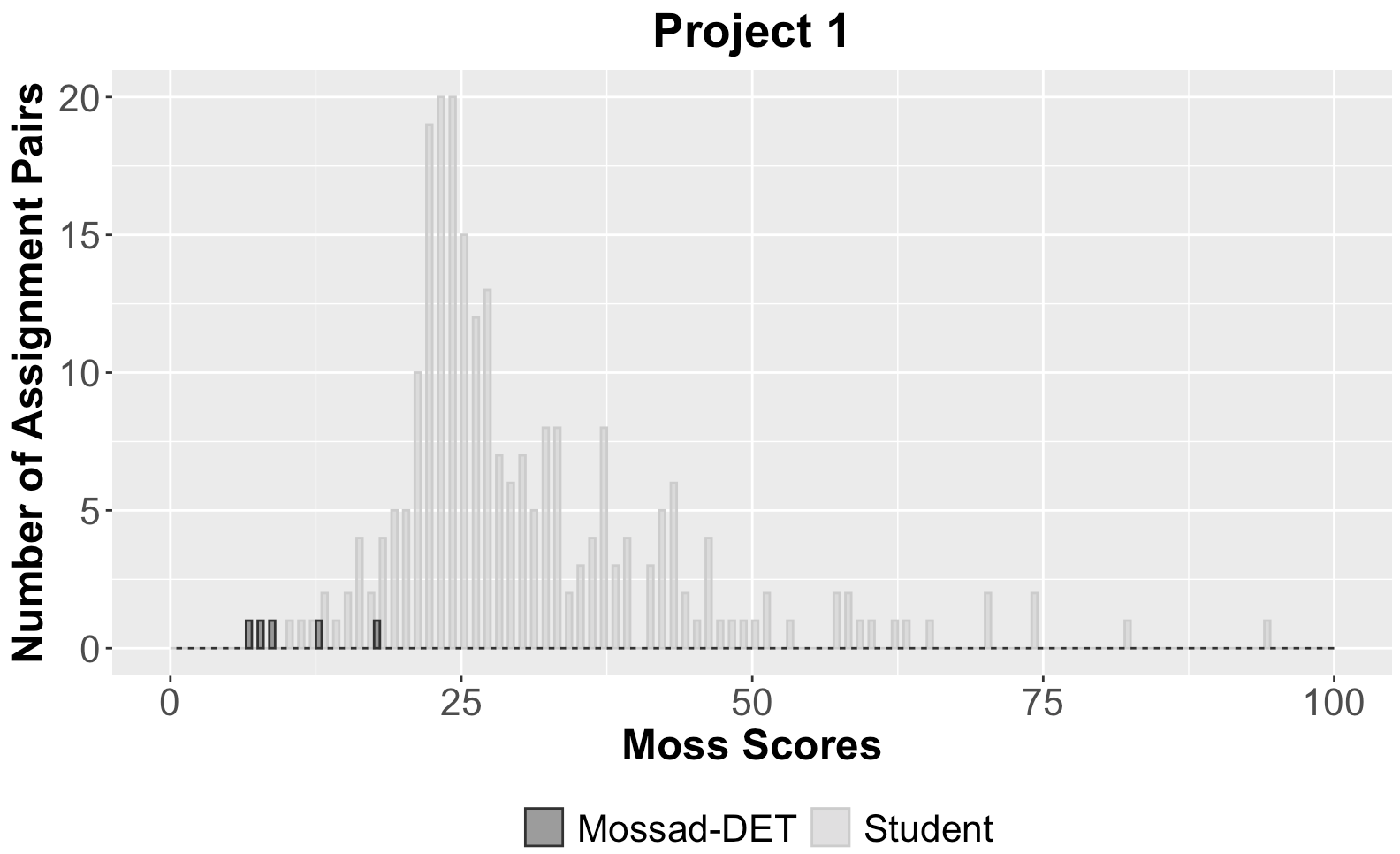}
          \caption{}
          \label{fig:m1lab4}
        \end{subfigure}\hfill%
        \begin{subfigure}[t]{0.33\linewidth}
          \includegraphics[width=0.99\textwidth]{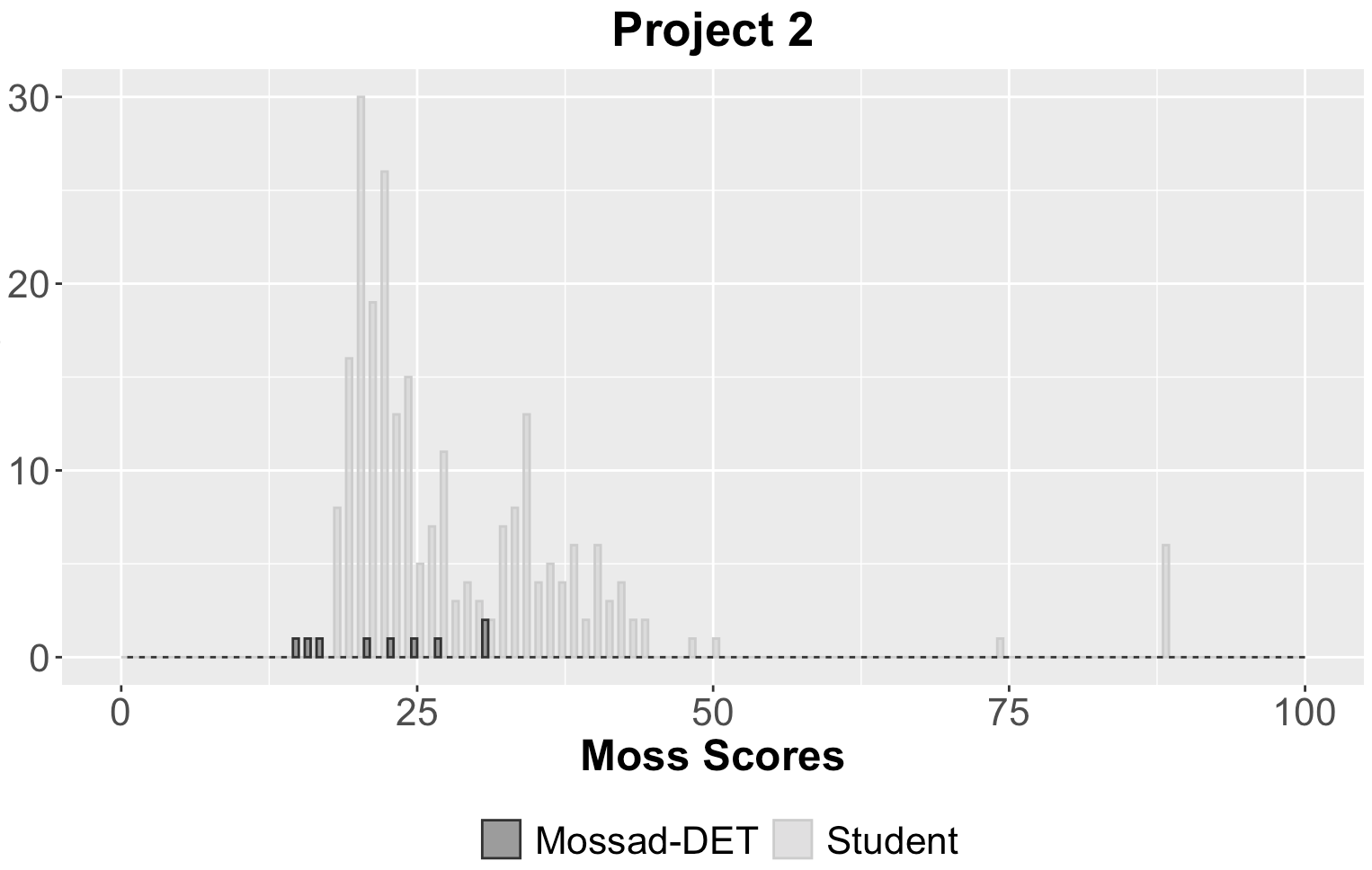}
          \caption{}
          \label{fig:m1lab7}
        \end{subfigure}\hfill%
        \begin{subfigure}[t]{0.33\linewidth}
          \centering\includegraphics[width=0.99\textwidth]{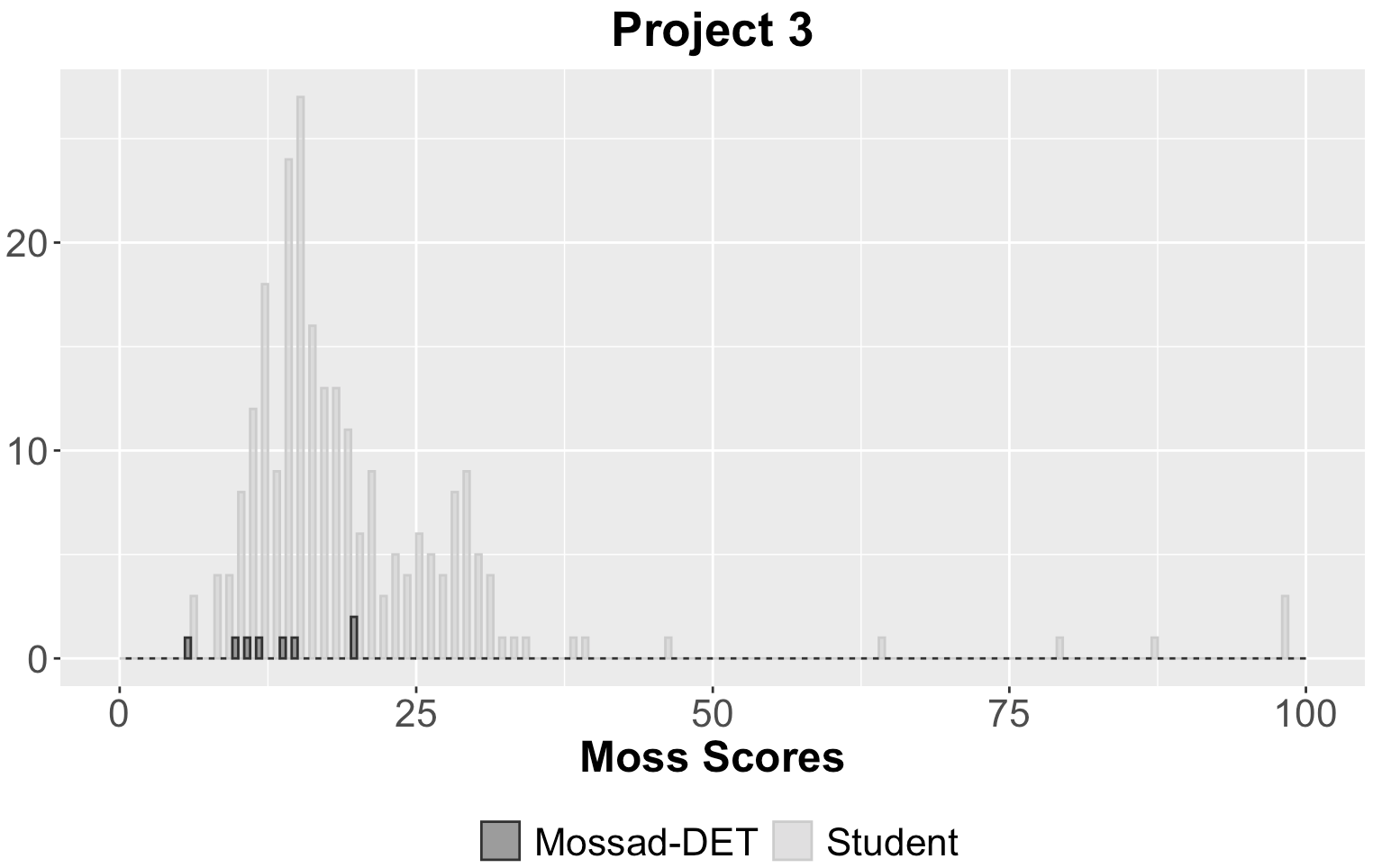}
          \caption{}
          \label{fig:m1lab8}
        \end{subfigure}
        \caption{\textbf{Ablation Study: \systemclone{} is also successful at defeating Moss's detection.} The dark gray bars denote a match between at least one \systemclone{} file, and the light gray bars denote purely authentic code matches. As Section~\ref{sec:ablation} describes, \systemclone{} is strictly less powerful than \systemfull{} and its deterministic nature makes mass plagiarism impossible.\label{fig:eval_m1}}
 
\end{figure*}

\begin{itemize}

  \item \textbf{Deterministic insertion of the same kind of
  statement.} \systemclone{} is an entirely deterministic variant that
  attempts to insert benign lines of code (fresh variable
  declarations) within each fingerprint
  window. Unlike \systemname{}, \systemclone{} always produces the
  same output from the same input program, running the risk of
  collision if assignments are plagiarized from the same input. This
  determinism also precludes mass plagiarism based on \systemclone{}.

  \item \textbf{Non-deterministic insertion of randomly-chosen
  statements.} \systemshatter{} is a non-deterministic variant that
  randomly selects locations within each fingerprint window, and
  inserts randomly chosen lines from the program. This non-determinism
  reduces the likelihood that two programs will collide, though we
  expect that its effectiveness for mass plagiarism would be limited.

\end{itemize}

\subsubsection*{Methodology}

To examine the effects of these variants, we use five randomly
selected files from each of our three datasets, and generate variants
with both \systemclone{} and \systemshatter{}. We separately augment
each dataset with the five corresponding plagiarized files and use
Moss to produce similarity scores for each dataset.

\subsubsection*{Fingerprint Windows}

To perform this ablation study, we need to know a single parameter:
the length of the fingerprint windows. As
Section~\ref{sec:hash_disruption} notes, this parameter is not
publicly available, and needed to be reverse-engineered. In fact, the
actual parameter we need is size of the fingerprint window in terms of
the number of lines of code, which we refer to as $w$. However,
different lines of code naturally lead to different numbers of
tokens.

We performed extensive experimentation across our suite of student
solutions and found that we could reliably perform hash disruption by
inserting a variable declaration between every three to four lines of
code. We therefore use $w=4$ lines of code as the window size for
these experiments.

\begin{figure*}[!t]
	\centering
        \begin{subfigure}[t]{0.34\linewidth}
          \includegraphics[width=0.99\textwidth]{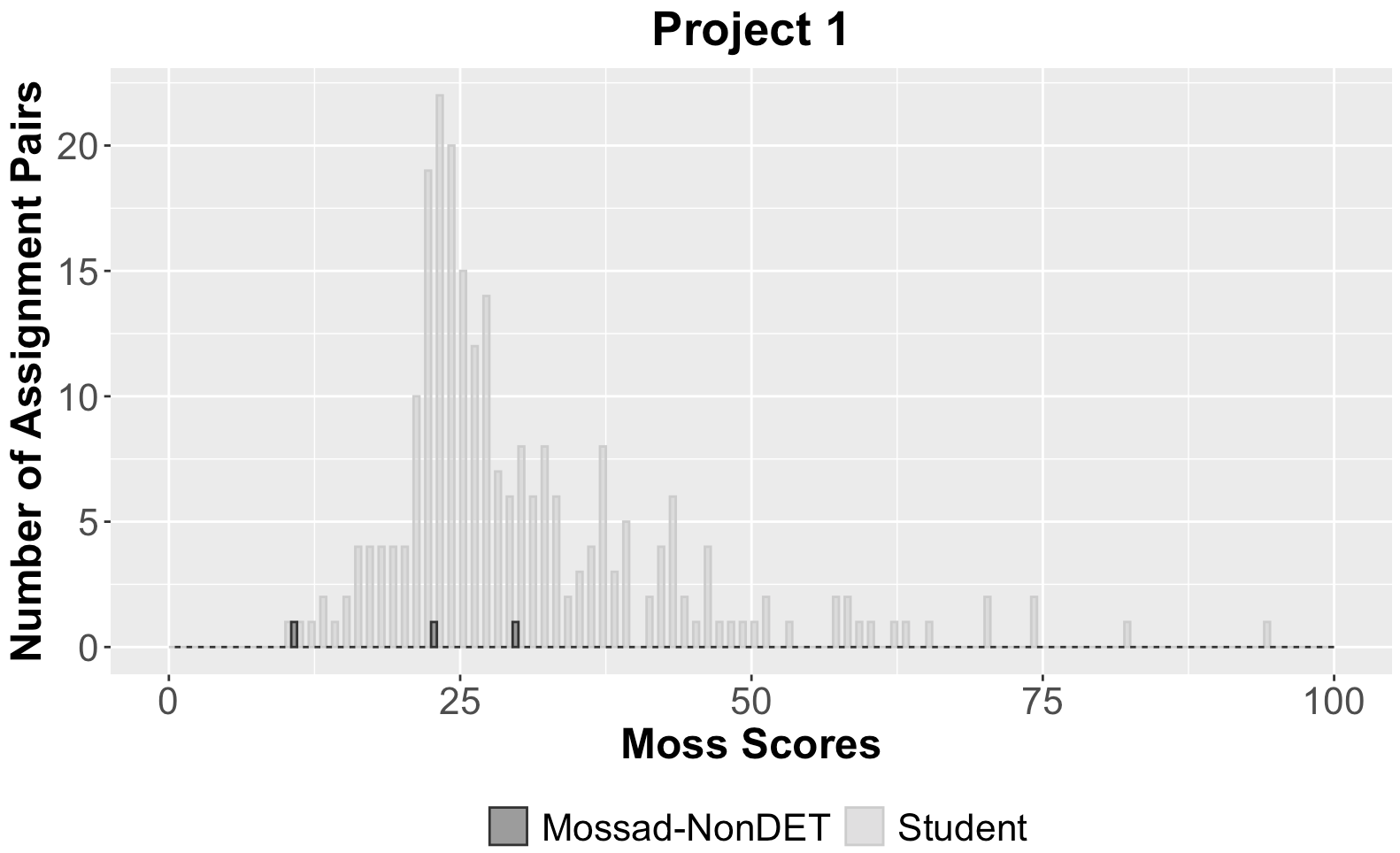}
          \caption{}
          \label{fig:m2lab4}
        \end{subfigure}\hfill%
        \begin{subfigure}[t]{0.33\linewidth}
          \includegraphics[width=0.99\textwidth]{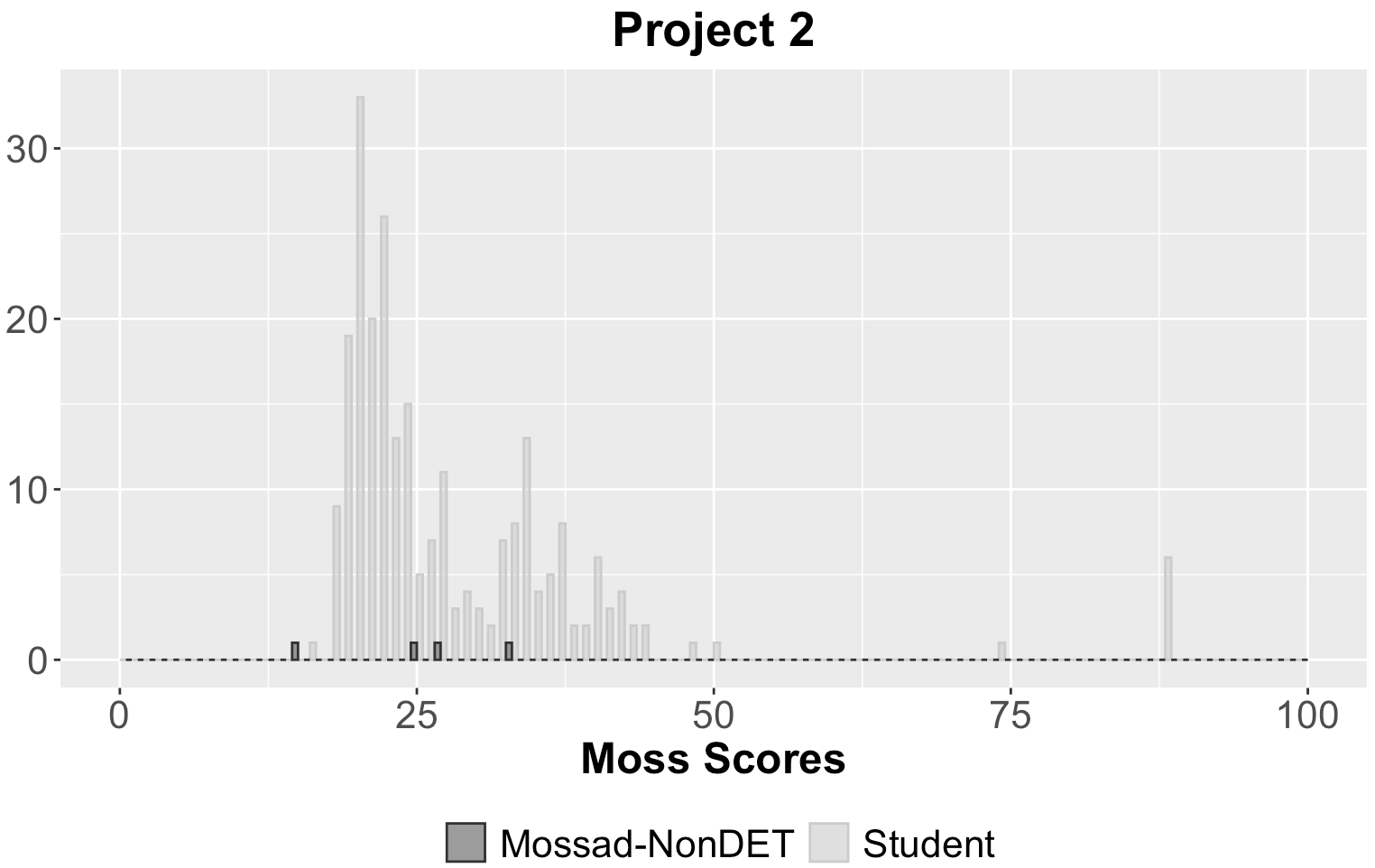}
          \caption{}
          \label{fig:m2lab7}
        \end{subfigure}\hfill%
        \begin{subfigure}[t]{0.33\linewidth}
          \centering\includegraphics[width=0.99\textwidth]{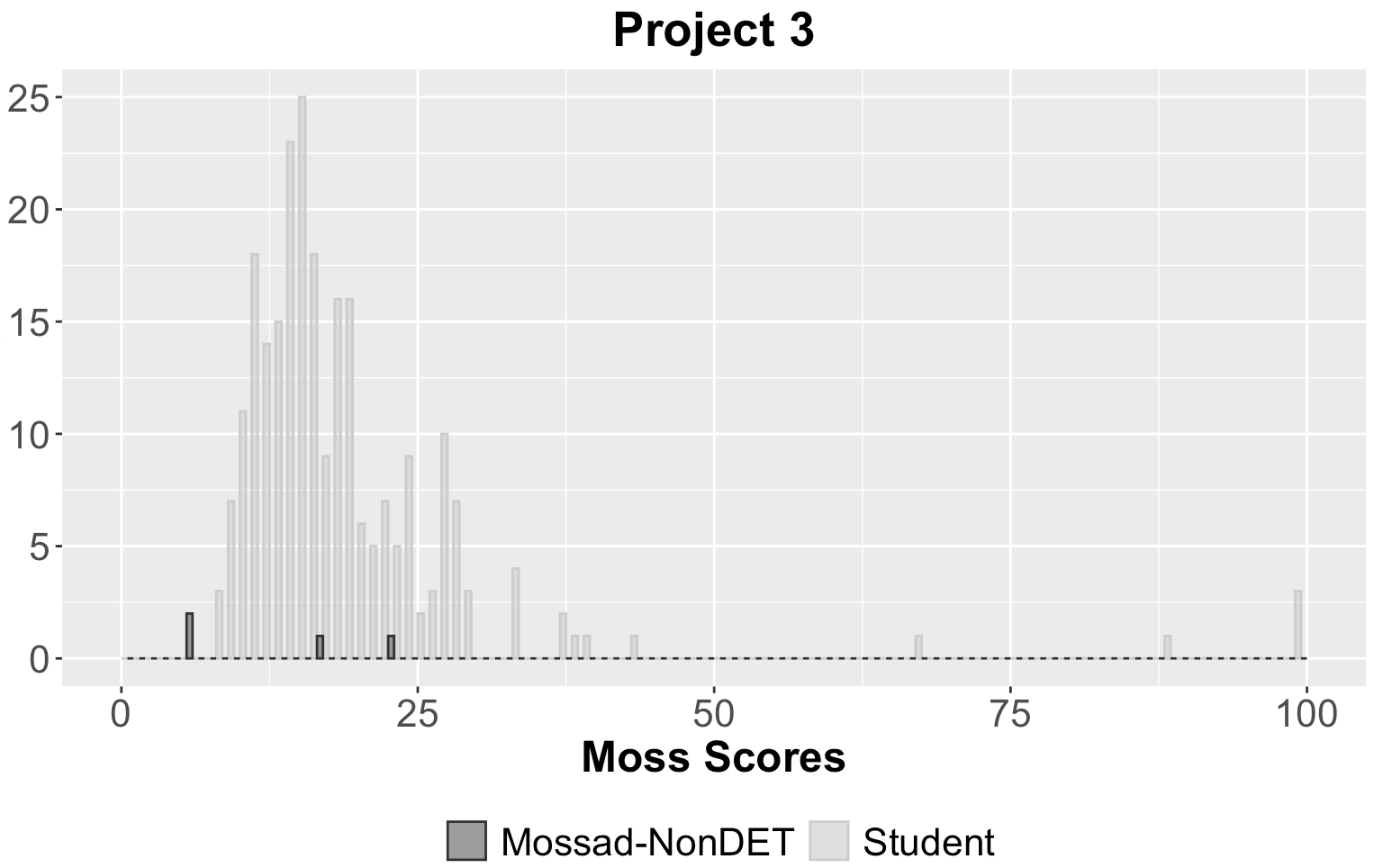}
          \caption{}
          \label{fig:m2lab8}
        \end{subfigure}
  \caption{\textbf{Ablation Study: \systemshatter{}'s non-deterministic approach also defeats Moss.} \systemshatter{} also occasionally leads to multiple variants producing high similarity scores with each other, but less frequently than its deterministic counterpart \systemclone{}.
  \label{fig:eval_m2}
  }
\end{figure*}

\subsubsection{\systemclone{}}
\label{sec:mossad_det}

We examine the effectiveness of \systemclone{} at producing code
variants that achieve low similarity scores when compared with the
original base file by using our system to perform its transformations on five
randomly selected files from each data set. After generating these
variants, we use Moss to score the similarity of each entire dataset.
Figure~\ref{fig:eval_m1} presents these results. The dark gray 
bars denote
comparisons that were made with at least one \systemclone{} variant;
most scored below the threshold of 25\%.

The deterministic nature of \systemclone{} introduces risk associated
with high Moss scores of multiple \systemclone{} variants. This risk
is visible in Figure~\ref{fig:m1lab7}: the dark gray bars that 
scored higher
than 25\% are actually comparisons made between two \systemclone{}
files. Since the additions that \systemclone{}  makes to an input file are
the same, it can cause Moss to detect additional, synthetic
similarity. Since multiple \systemclone{} variants are all generated
from the same base file, two attempts to plagiarize from the same
source code will result in Moss scores of 100\%.

\subsubsection{\systemshatter{}}
\label{sec:mossad_nondet}

The \systemshatter{} variant works by iterating through all non-overlapping windows of
size $w$ in consecutive order, trying to insert a randomly selected
line from the source program into a randomly selected position within
that window. If the insertion produces a compilation error or the
object code does not match the original, the system will repeat this
process $wlog(w)$ times. If all attempts fail, \systemshatter{} moves 
on to the next window.

By using the source file itself as the source of code to
insert, \systemshatter{} adds greater non-determinism that lowers the
probability that the same statement was added to the same window. When
combined with the nondeterminism introduced by randomly choosing an
insertion location within each window, we expect \systemshatter{} to
lower the number of matches of hashes across multiple variants
generated from the same base file. Figure~\ref{fig:eval_m2} presents
the results. 

\systemshatter{} consistently scores below the threshold similarity
score of 25\% in each project; however, for Projects 1 and 2
the scores center around 25\%, with each score between 6\% and 30\%.
Since \systemshatter{} is not fully nondeterministic, the resulting
Moss scores of its outputs are shown to occasionally under-perform
\systemfull{}; this is because it has limited areas in which to insert
mutations, resulting in decreased chances of generating
a mutation that is successful in compilation \textit{and} in
object code comparison.

\begin{figure}[!t]
    \centering
    \includegraphics[width=80mm]{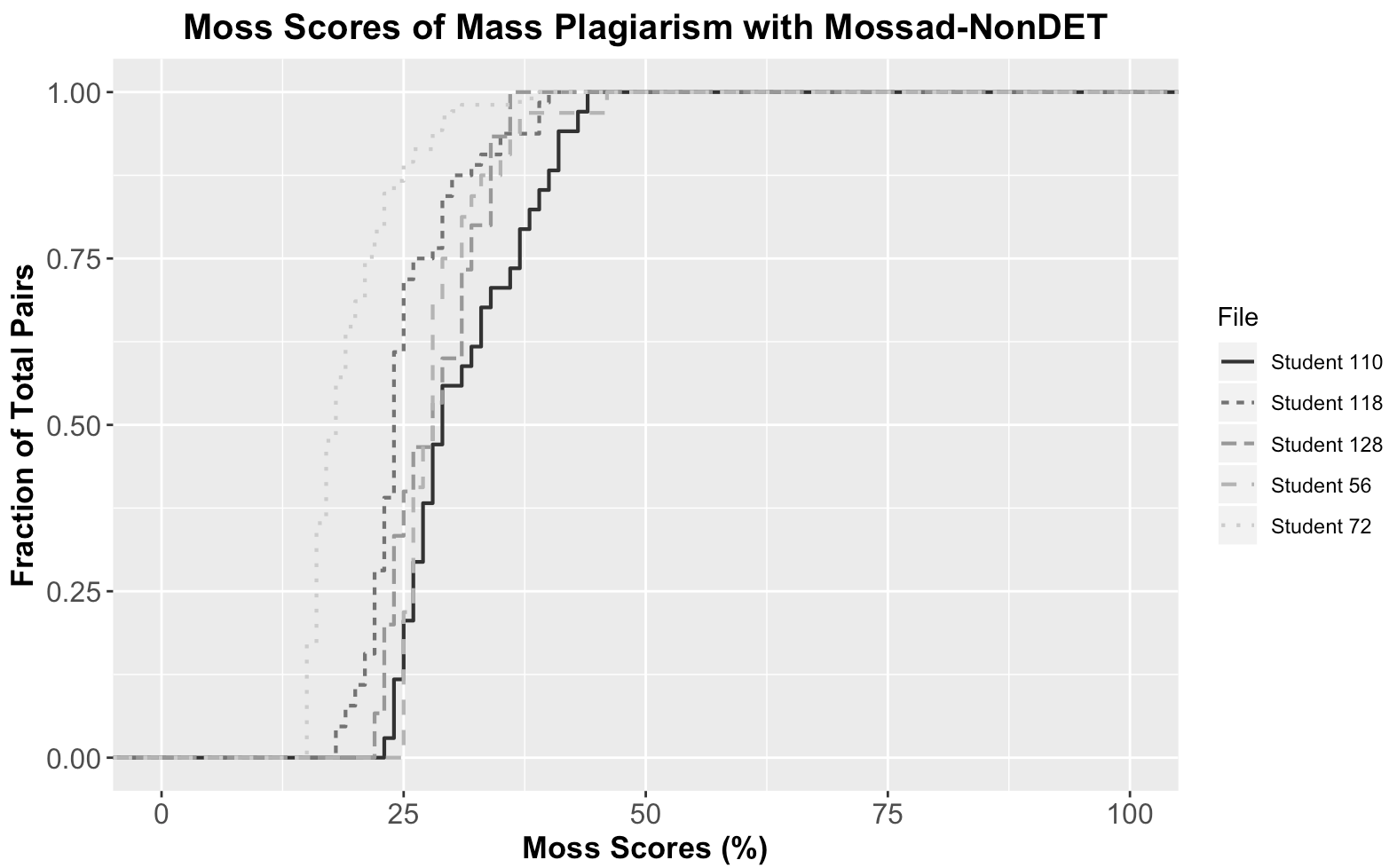}
    \caption{\textbf{Mass plagiarism with \systemshatter{}
        frequently yields suspicious Moss scores.}  Unlike
      \systemfull{}, \systemshatter{} produces variants with
      suspicious scores in some cases roughly 80\% of the time.}
    \label{fig:mass_shatter}
\end{figure}

Because \systemshatter{} is non-deterministic, it should in principle
be applicable to mass plagiarism. We perform the same experiment
described in Section~\ref{sec:mossad_mass_plagiarism}. As
Figure~\ref{fig:mass_shatter} shows, it can produce variants with
suspicious scores ($>25\%$) as much as 80\% of the time. That said,
90\% of the pairs of files score a similarity score below 40\%. While
this similarity is generally low, it significantly underperforms
\systemfull{} in this scenario (maximum similarity: 27\%).

\subsubsection{Discussion}
\label{sec:ablation_discussion}
The major difference between \systemname{} and the ablated versions is
the level of determinism. \systemname{} is fully nondeterministic;
that is, it is free to place its inserted lines anywhere, giving it a
higher likelihood of disrupting hash windows. The two ablated versions
can at most insert one line of code every so many lines of code. Since
lines of code may not actually align with windows (which are actually
in units of tokens, not lines of code), the ablated versions may fail
to disrupt hashes that \systemname{} can disrupt.

\begin{shaded}
\textbf{Ablation Study}: \emph{While \systemclone{} and \systemshatter{} are generally effective at undermining Moss, \systemfull{} dominates both of them, especially for mass plagiarism.}
\end{shaded}

\subsection{Threats to Validity}
\label{sec:threats}

In this section, we examine the threats to validity of the evaluation of
\systemname{}.

\subsubsection{\systemname{} File Sizes}

Due to the nature of code insertion, \systemname{} files are always
longer than the base file, though the current implementation limits
the number of consecutive insertions to ensure that files do not grow
too large. For the datasets used throughout the experiments described
earlier in this section, the number of lines of code ranges from 10 to
100 (which in and of itself presents another threat to validity); longer
code produced by \systemname{} attacks would not necessarily be an
easily-identifiable pattern since this (legitimate) variance is
already present in the dataset. However, the length of \systemname{}
files may be a characteristic that could identify suspiciousness for
other datasets in which we have not examined.

\subsubsection{Populations}

The two surveys used to inform the evaluation have very different
results; as previously noted, the surveys themselves were
different. The Twitter-only survey was brief and could be answered
with a single click, while the second survey required data entry into
a Google Form. The latter may have been a barrier to
participation. Although the two populations disclosed varying degrees
of manual inspection, the user study presented in
RQ2~\ref{sec:neena_experiment} suggests that even when code is
manually inspected, \systemname{}-generated code escapes detection.

For the user study presented in RQ2~\ref{sec:neena_experiment}, the
population is taken from a single institution, an R1 public
university, and may not be representative. While it is a reasonable
size (N=30), a larger sample size may result in different
conclusions. Additionally, changes to the study, such as providing a
different rubric or grading metric or removing the time limit and
letting graders choose the amount of time to spend on manual
inspection, could lead to different results.

\subsubsection{Moss Latency}

For the performance evaluation of \systemname{} in 
RQ5~\ref{sec:mossad_performance}, the performance of \systemname{} is heavily
dependent on the latency and thresholds of the deployed Moss system, and is 
subject to variance due to dynamic load. The point of this experiment was to 
examine \systemname{} performance in the real world. These experiments could inadvertently 
have used Moss during times of low load; higher load could increase the 
average time taken for \systemname{} and consequently could have had an impact on 
the real world results.

\section{Obfuscation}
\label{sec:obfuscation}

\begin{figure}[!t]
  \centering
  \begin{subfigure}[!b]{0.45\textwidth}
  \centering
  \hspace{-1em}
  \includegraphics[height=30mm]{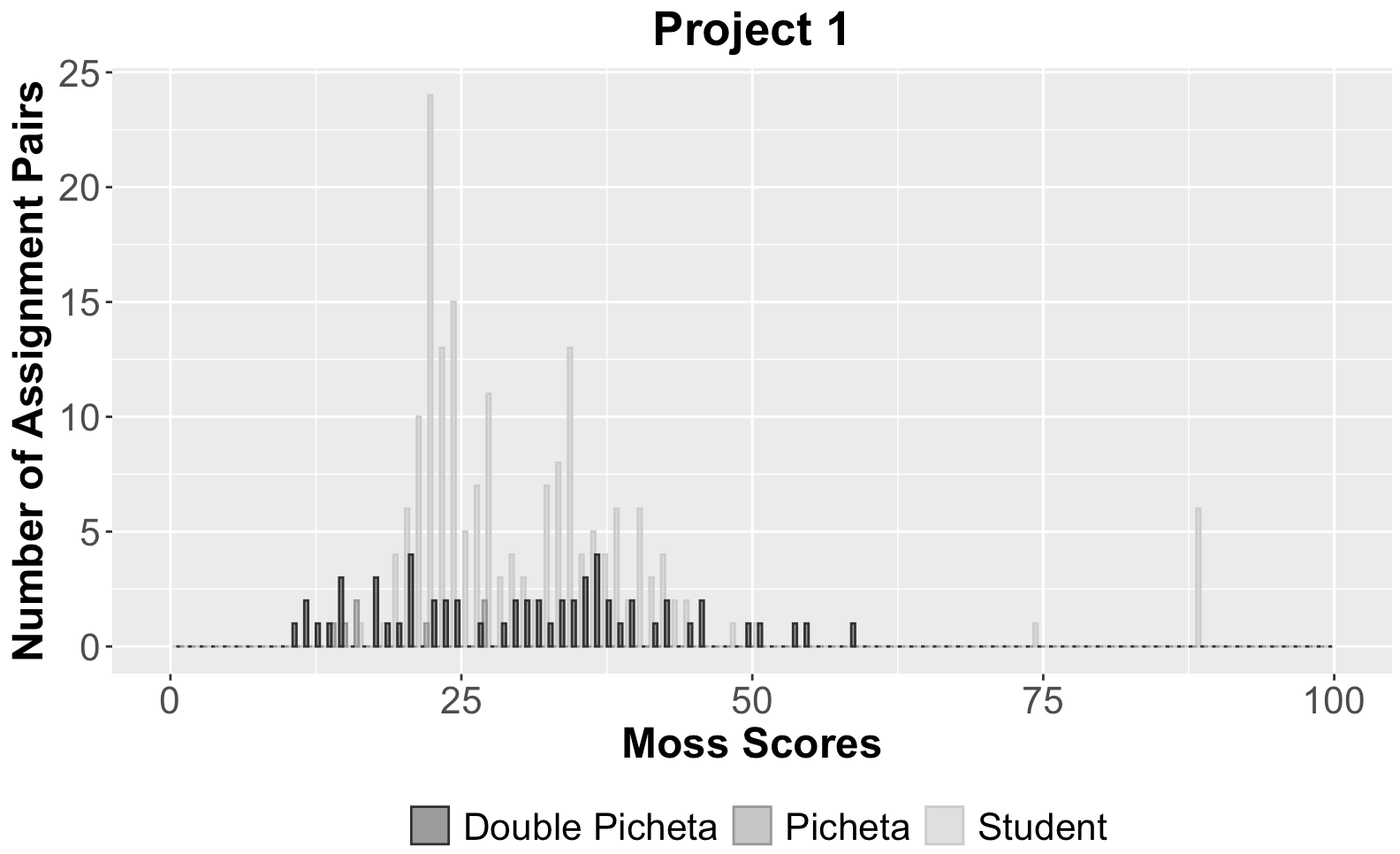}
  \caption{\textbf{Obfuscation can defeat Moss's similarity detection algorithms, though not reliably.}
  \label{fig:picheta_scores}
  }
  \end{subfigure}
  \hspace{1em}
  \begin{subfigure}[!b]{0.45\textwidth}
  \centering
  \includegraphics[height=30mm]{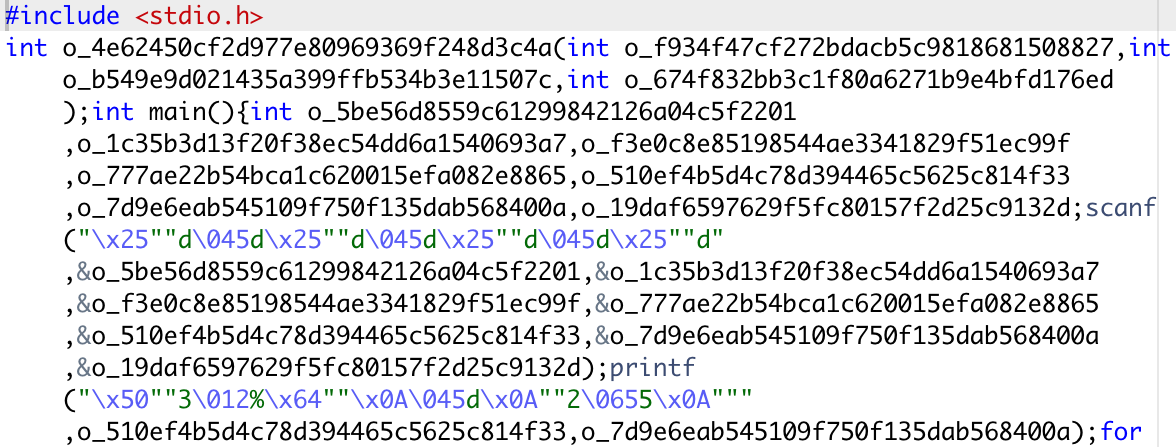}
  \caption{\textbf{Obfuscation produces highly unreadable code.}\label{fig:obfuscated_code}}
  \end{subfigure}
  \caption{\textbf{Obfuscators can defeat Moss's similarity detection
algorithms, but at a cost.}  Picheta is the most effective of the
obfuscators we evaluated at producing low similarity scores. However,
it is not consistently reliable, frequently producing highly
suspicious matches. It suffers from two other major drawbacks: (1) it
is deterministic, so code plagiarized from the same source will yield
a high match, and (2) it produces obviously-obfuscated
code.\label{fig:picheta}}
\end{figure}

This section explores the possibility of an alternative approach
to \systemfull{}: code obfuscation.  Intuitively, this approach may
seem appealing. By definition, the goal of an obfuscator is to
transform a program so that it is not as legible as the original
program. This effect appears to hold promise as a means of thwarting
software plagiarism detectors. However, we find this is not the case.

We tested a number of publicly-available obfuscators. We
found that many of the obfuscations are ones that Moss trivially
defeats, such as whitespace removal. We discuss here the obfuscator
that was most successful at defeating Moss's detection; however, as we
describe below, we find that it suffers from several significant
drawbacks that make it ineffective as an attack on software plagiarism
detection.

The obfuscator we evaluate here is an online obfuscator for C/C++,
whose stated intended use is to protect against software piracy, reverse
engineering, and
tampering~\cite{picheta2020code,picheta2020page}. This obfuscator has
no name; we refer to it here as \emph{Picheta} after its
author. Picheta performs three major transformations:
\emph{data}, where data is transformed into different radices or
turned into a static expression; \emph{lexical}, where all identifiers
are hashed; and \emph{control}, where the semantics of the program are
slightly modified, including dead code insertion. Of the three
obfuscators we found that were at least moderately successful in
undermining detection by Moss (Stunnix~\cite{stunnix}, Tigress~\cite{tigress}, and Picheta), we
found that Picheta was the most successful in producing low Moss
similarity scores.

\subsection{Evaluation}
\label{sec:obfuscator_eval}

Using Picheta to plagiarize source files from our datasets, we
examined its effectiveness as a means to avoid plagiarism
detection. For this experiment, we randomly selected 10 files from the
first dataset as inputs to Picheta. The resulting ten obfuscated files
were added back to the original set of assignments to act as
plagiarized files; the resulting corpus was compared using Moss.

Figure~\ref{fig:picheta_scores} graphs the Moss scores of the top 250
highest-scoring pairs of files. We also performed this experiment
using the two additional data sets; the results were similar, so we
omit those graphs. We find that Picheta is moderately effective, but
roughly 25\% of the time, it produces highly suspicious outputs, with
resulting Moss similarity scores substantially higher than our suspiciousness
threshold (40\%--98\%).

Like all other obfuscators we examined, Picheta is also deterministic.
Thus, plagiarism from the same base file using Picheta would result in
the exact same files, leading to likely discovery.  Interestingly, as
shown in Figure~\ref{fig:picheta_scores} (denoted by \textit{Double
Picheta}) many of the matches with high similarity scores are actually
artifacts of using the obfuscator itself, further limiting its
effectiveness, as obfuscated code is self-similar even when the input
programs differ. Beyond these drawbacks, the code produced by the
obfuscator is immediately suspicious, as shown in
Figure~\ref{fig:obfuscated_code}.

\section{Countermeasures}
\label{sec:countermeasures}
This section describes a limited countermeasure to
the \systemname{} line of attacks.
The current prototype implementation of \systemfull{} relies on
introducing semantics-preserving code in an extremely conservative manner:
directly comparing generated object files. The code that \systemfull{}
introduces is unreachable, redundant, or dead.

We therefore evaluated the effectiveness of the following
countermeasure: first, compile all code with a high optimization level
(e.g., \texttt{-O3}) and output \emph{assembly code} (via
the \texttt{-S} flag). Second, submit the assembly code to Moss,
indicating it should use one of the supported ISAs as its input
language (Moss currently supports both x86 and MIPS). We found that
for each project, this countermeasure undoes the effect of \systemfull{}
by optimizing away its code insertions.

Unfortunately, this countermeasure suffers from several
limitations. First, this approach requires that instructors manually
map back matched assembly code fragments to the original source files,
which is inconvenient at best.  Second, and more seriously, generated
assembly code---especially at high optimization levels---results in
spurious matches due to compiler-generated artifacts like function
prologues and epilogues.  Third, it only works when optimizing
compilers are available, and where it is possible to emit ``clean''
assembly code (e.g., without inlined calls to parts of a runtime
library).  It is not obvious if this is even possible currently with
existing JIT compilers for languages like Python, Ruby, or JavaScript.

The results of this paper also suggest that more extensive code
review, although costly in terms of human effort, could be an
effective countermeasure to the \systemname{} line of attacks;
empirical evaluation of this countermeasure is future
work. Additionally, we hypothesize that integrating version control
into the process of code assignment submission and subsequent grading
could also mitigate the effects of \systemname{}; exploring this is
also future work. We believe that both of these potential
countermeasures have additional positive benefits beyond plagiarism
detection: code review and version control are both useful skills for
computer scientists and programmers, and are a way to give meaningful
feedback to students.

\section{Related Work}
\label{sec:relatedwork}
\label{sec:related_work}

This section describes other software plagiarism approaches not
discussed earlier in the paper.  We note that, in an effort to further
evaluate \systemname{}'s effectiveness beyond Moss, JPlag,
Sherlock, and Fett, we contacted the authors of these papers; all authors
responded that their tools are not available, precluding a direct
empirical comparison. 

\paragraph{Static techniques:} To combat students shuffling independent code segments to thwart plagiarism
detection, Wise presents a string-similarity approach using
Running-Karp-Rabin Greedy-String-Tiling, which can detect transposed 
subsequences in many source languages~\cite{Wise:1996:YID:236452.236525};
this is the technique employed by Sherlock.
Liu et al.\ and Chae et al.\ present
graph-based approaches to detecting software 
plagiarism~\cite{Liu:2006:GDS:1150402.1150522,Chae:2013:SPD:2505515.2507848}.
In the first case, plagiarism is detected by mining
program dependence graphs. This method of plagiarism detection is effective because
program dependence graphs are generally invariant in the face of
plagiarism~\cite{Liu:2006:GDS:1150402.1150522}. However, extending
\systemname{} to adding unnecessary control flow
to a plagiarized file could thwart this detection technique. Chae et
al.\ also construct a control flow graph of a program; they then
perform a random walk to compute the importance of each node, and
finally generate a single score vector of the graph for
comparison~\cite{Chae:2013:SPD:2505515.2507848}. Son et al.\ uses the
structural information of the program (obtained via parsing) to
determine similarity~\cite{Son2006}. Luo et al.\ propose a
binary-oriented program similarity method based on longest common
subsequences of semantically equivalent basic
blocks~\cite{Luo:2014:SOB:2635868.2635900}. 

\paragraph{Dynamic techniques:}
Jhi et al.\ introduce value-based program characterization
leveraging invariant values, in an effort to create a program
plagiarism detection algorithm that is resilient to control flow and
data obfuscation techniques~\cite{Jhi:2011:VPC:1985793.1985899}. 

\paragraph{Code Clone Detection:}
Another potential method for identifying possible plagiarism attempts is via
code clone detection. We evaluated SourcererCC~\cite{1512.06448}, a
recent code clone detector used to map code duplicates on
Github~\cite{Lopes:2017:DMC:3152284.3133908}, in an effort to defeat
\systemname{}. SourcererCC works at two granularities:
file-level and block-level. We examined the output of running
SourcererCC at the file-level granularity on a corpus of one of our C
datasets and various \systemname{} variants. SourcererCC returned
numerous near-miss clones, which the authors describe as clones
``where minor to significant editing activities might have taken place
in the copy/pasted fragments''~\cite{1512.06448}. However, SourcererCC
returned nearly every file from the corpus as near-misses (these are
almost all false positives). We do not believe that file-level
operates at a sufficiently fine granularity for detecting code
plagiarism that has been disguised using \systemname{}. SourcererCC's
block-level granularity only supports Python and Java; we were unable
to test if this granularity is sufficiently fine to detect disguised
plagiarism using \systemname{} with our projects, which are written in C.


\section{Conclusion}
\label{sec:conclusion}
This paper presents \systemname{}, a fully-automated program
transformation system that defeats software plagiarism detection. We
demonstrate its efficacy on Moss, the most effective and most
widely-used such detector, as well as on JPlag. \systemname{}
transforms source programs into one or even dozens of variants that
all escape their detection algorithms while maintaining
readability. Because \systemname{} directly strikes at the algorithmic
underpinnings of Moss and similar systems, effectively coping with
these will require innovation in software plagiarism detection.

\textit{Acknowledgments.} We thank Don Porter for providing our anonymized C code datasets.
We also thank Neena Thota for providing time and space in her classroom for our case study,
and her students for volunteering to be a part of the case study. This material is based upon
work supported by the National Science Foundation under Grants No. CCF-1439008 and CCF-1617892.

\clearpage

{\normalsize
\bibliography{mossad}}

\clearpage

\appendix
\section{Survey Instrument}

\begin{figure}[H]
	\centering
	\includegraphics[scale=0.5]{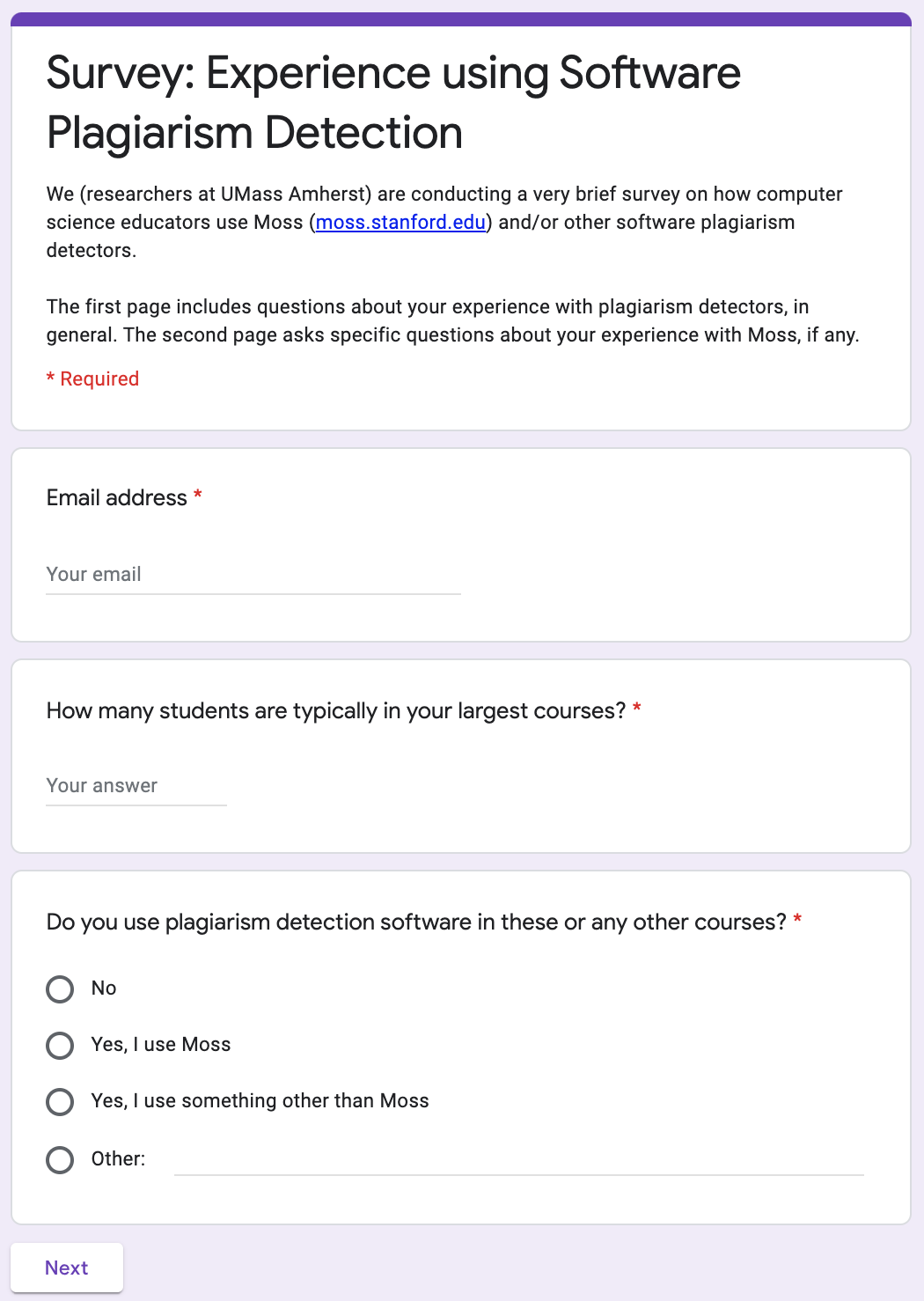}
	\caption{Survey respondents were asked to verify their eligibility by 
	providing their institutional email address. Personally identifiable information was removed from those responses. Additionally, respondents were provided with our IRB approval with this survey.}
	\label{fig:survey1}
\end{figure}

\begin{figure}[tb]
	\centering
	\includegraphics[scale=0.25]{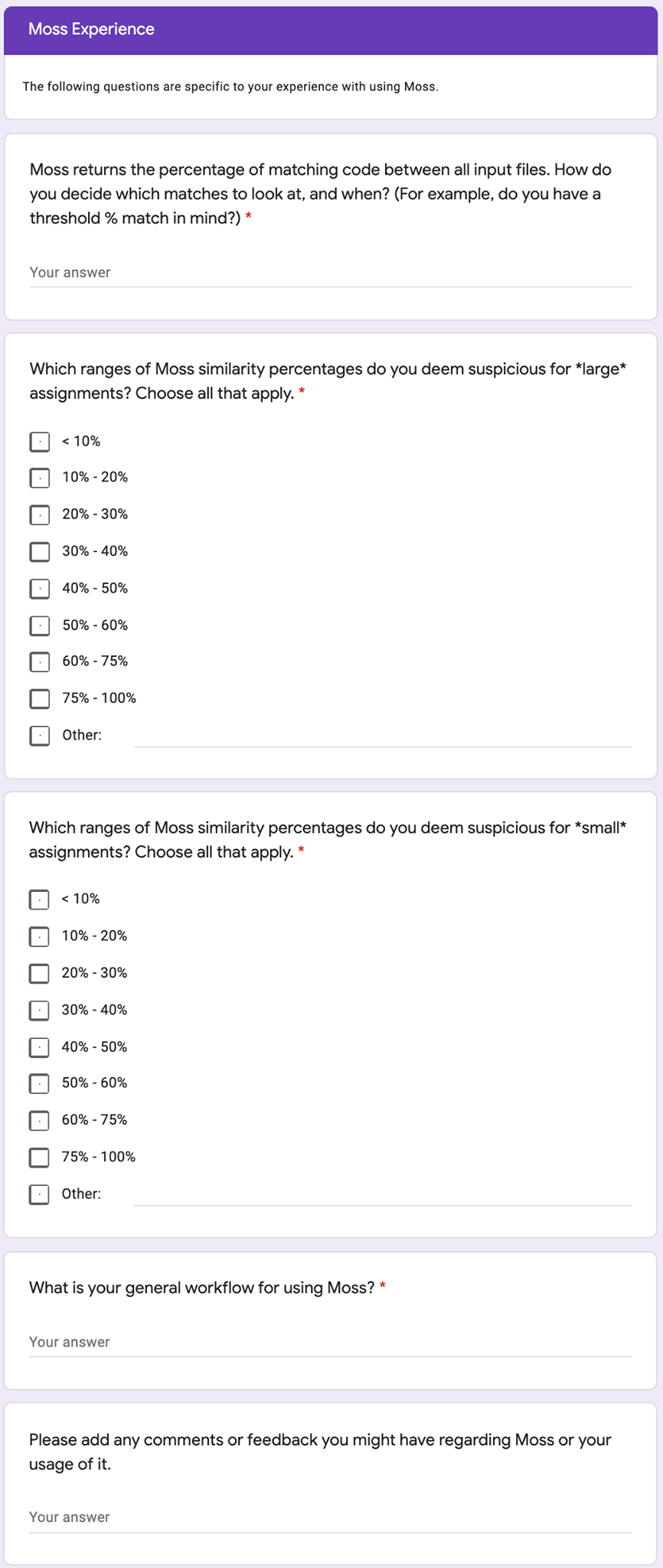}
	\caption{If a respondent reported that they used Moss in their classrooms, they were instructed to describe their workflow and report which Moss percentages were suspicious. Lastly, survey respondents were given space to report any additional information they wanted to share.}
	\label{fig:survey2}
\end{figure}

\end{document}